\documentclass[fleqn,10pt]{wlscirep}

\usepackage[utf8]{inputenc}
\usepackage{graphicx,epsfig}
\usepackage{times}
\usepackage{mathtools,amssymb,amsmath,amsfonts,multirow,rotate,color}
\usepackage[caption=false]{subfig}
\usepackage{color}
\usepackage{soul}
\usepackage{tikz}
\usepackage{verbatim}
\usepackage{pgfplots}
\usepackage{appendix}
\usepackage{mathrsfs}
\usepackage{wasysym}
\usepackage{stmaryrd}
\usepackage{chngcntr}
\usepackage{algpseudocode,algorithm}

\usetikzlibrary{patterns}
\usetikzlibrary{positioning}

\pgfplotsset{compat=1.15}

\title{The Weighted Bitcoin Lightning Network}

\author[1,2,3]{Jian-Hong Lin}
\author[4]{Emiliano Marchese}
\author[1,2]{Claudio J. Tessone}
\author[4,5]{Tiziano Squartini}

\affil[1]{Blockchain and Distributed Ledger Technologies, Institute of Informatics, University of Z\"urich, Andreasstrasse 15, CH-8050 Z\"urich (Switzerland)}
\affil[2]{UZH Blockchain Center, University of Z\"urich, Andreasstrasse 15, CH-8050 Z\"urich (Switzerland)}
\affil[3]{ETH Z\"urich, Department of Management, Technology and Economics, Scheuchzerstrasse 7, CH-8092 Z\"urich (Switzerland)}
\affil[4]{IMT School for Advanced Studies Lucca, Piazza San Francesco 19, I-55100 Lucca (Italy)\thanks{tiziano.squartini@imtlucca.it}}
\affil[5]{Institute for Advanced Study, University of Amsterdam, Oude Turfmarkt 145, NL-1012 GC Amsterdam (The Netherlands)}

\begin{abstract}
The Bitcoin Lightning Network (BLN) was launched in 2018 to scale up the number of transactions between Bitcoin owners. Although several contributions concerning the analysis of the BLN binary structure have recently appeared in the literature, the properties of its weighted counterpart are still largely unknown. The present contribution aims at filling this gap, by considering the Bitcoin Lightning Network over a period of 18 months, ranging from $12^\text{th}$ January 2018 to $17^\text{th}$ July 2019, and focusing on its weighted, undirected, daily snapshot representation - each weight representing the total capacity of the channels the two involved nodes have established on a given temporal snapshot. As the study of the BLN weighted structural properties reveals, it is becoming increasingly `centralised' at different levels, just as its binary counterpart: 1) the Nakamoto coefficient shows that the percentage of nodes whose degrees/strengths `enclose' the $51\%$ of the total number of links/total weight is rapidly decreasing; 2) the Gini coefficient confirms that several weighted centrality measures are becoming increasingly unevenly distributed; 3) the weighted BLN topology is becoming increasingly compatible with a core-periphery structure, with the largest nodes `by strength' constituting the core of such a network, whose size keeps shrinking as the BLN evolves. Further inspection of the resilience of the weighted BLN shows that removing such hubs leads to the network fragmentation into many components, an evidence indicating potential security threats - as the ones represented by the so called `split attacks'.
\end{abstract}

\begin{document}

\maketitle

\thispagestyle{empty}

\section{Introduction}

The Bitcoin Lightning Network (BLN) \cite{poon2016bitcoin} represents an attempt to overcome one of the main limitations of the Bitcoin technological design, i.e. \emph{scalability}: at the moment, only a limited amount of transactions per second, whose number is proportional to the size of blocks and their release frequency, can be processed by Bitcoin, a major shortcoming preventing the adoption of this payment system at a global scale - especially when considering that classic payment mechanisms are able to achieve tens of thousands of transactions per second. Increasing the size of the blocks has been proposed as a solution; implementing this choice, however, would require 1) a larger validation time, 2) a larger storage capability and 3) larger bandwidth costs, hence favoring a more centralised validation process: in fact, fewer entities would become able to validate the new blocks, thus making the system as a whole more prone to faults and attacks.

\begin{figure}[t!]
\begin{center}
\includegraphics[width=\textwidth]{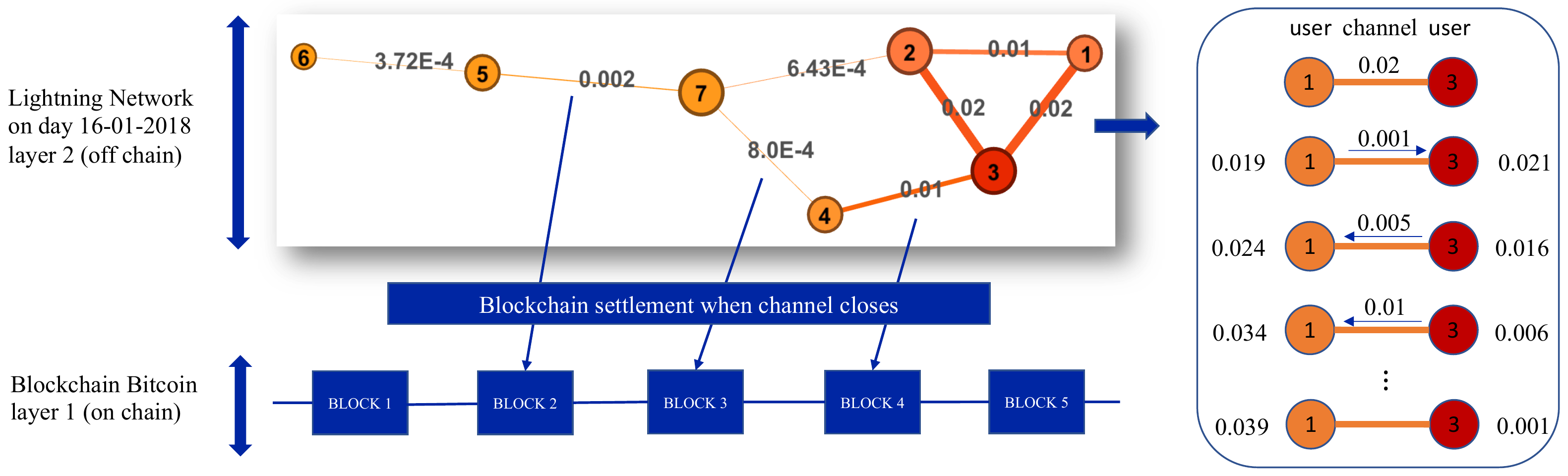}
\caption{Pictorial representation of the BLN functioning. The network structure refers to day 16-01-2018, i.e. when the network was constituted by seven nodes (representing users) and eight links (representing open channels). The size of each node is proportional to its degree (i.e. the bigger the node, the larger its degree) while the color of each node is proportional to its strength (i.e. the darker the node, the larger its strength). The width and color of each link are proportional to its weight, i.e. the capacity of the channel it depicts.}
\label{fig0}
\end{center}
\end{figure}

Developers have tried to break the trade-off between block size and centralisation by proposing to process transactions off-chain, i.e. by means of a `Layer 2' protocol that can operate on top of blockchain-based cryptocurrencies such as Bitcoin: nowadays, such a protocol is known with the name of \emph{Bitcoin Lightning Network} (BLN) and works by creating payment channels across which any two users can exchange money without having the data related to their transactions burdening the entire blockchain (see fig. \ref{fig0}).

The BLN has recently raised a lot of interest: analyses concerning the BLN functioning have found it to allow for a large number of transactions to be processed although rewarding fees are not high enough to promote a large participation in the mining process \cite{branzei2017charge}; Beres et al. \cite{beres2021cryptoeconomic} analysed transaction fees and privacy provisions, finding that the participation of most nodes is, in fact, `economically irrational'. 

On the other hand, analyses concerning the BLN network structure have found it to be characterised by a scale-free topology \cite{lee2020robustness}; besides, Lin et al. \cite{lin2020lightning} and Martinazzi et al. \cite{martinazzi2020evolving} considered the evolution of the former and found it to have become increasingly centralised at different levels. Both features - common to several cryptocurrency transaction networks such as Bitcoin, Bitcoin Cash, Dash, Dogecoin, Ethereum, Feathercoin, Litecoin, Monacoin and Z-Cash \cite{motamed2019quantitative,campajola2022evolution} - may have undesirable consequences such as causing a considerable fraction of payments to be easily de-anonymizable \cite{beres2021cryptoeconomic} and making it prone to channel exhaustion or attacks aimed at isolating nodes (thus, compromising their reachability, the payment success ratio, etc.) \cite{guo2019measurement,rohrer2019discharged}; a similar conclusion is reached in \cite{mizrahi2021congestion}, where the authors analyse the robustness of the BLN against three different types of attacks (locking channels, disconnecting pairs of nodes and isolating hubs) and find it to be disruptable at a relatively low cost; still, Conoscenti et al. \cite{conoscenti2019hubs} have suggested the BLN to be resilient against the removal of nodes that do not have a significant influence on the probability of success of a payment. More on the modeling side, Bartolucci et al. \cite{bartolucci2020percolation} proposed a fitness-based network model for the emergence of the BLN - a proposal that is reminiscent of the `fittest-gets-richer' principle, suggested by Javarone et al. \cite{javarone2018bitcoin}, as the mechanism driving the evolution of the global structure of the Bitcoin network and the Bitcoin Cash network.

For a summary of the existing literature on cryptocurrency transaction networks see Wu et al. \cite{wu2021analysis}.

\section{Data}

Most of the aforementioned contributions have just focused on the analysis of the BLN binary structure, leaving its weighted counterpart largely unexplored \cite{lin2020lightning}. With the present paper we aim at filling this gap, by studying the weighted properties of the BLN daily snapshot representation, at the micro-, meso- and macro-scale, across a period of 18 months, i.e. from $12^\text{th}$ January 2018 to $17^\text{th}$ July 2019.

In order to properly understand the structural details of the aforementioned representation, let us briefly sum up how the BLN works. Payments in the BLN are \emph{source-routed} and \emph{onion-routed} - features that are there to enhance the users privacy: while `source-routed' means that the source node is the one responsible for finding a route connecting itself to the final one, `onion-routed' means that the intermediate nodes only know the ones lying `one step-before' and `one step-after' them, in the payment route.

In order to pre-compute the entire payment route, the sender must have a reasonably up-to-date view of the network topology. Nodes in the BLN regularly broadcast information about the channels they participate in: such a mechanism, called \emph{gossip}, allows other nodes to keep their view of the network topology up-to-date.

The BLN topology can be visualised by means of the the so-called \emph{routing table}. For this paper, we took a snapshot of the routing table every 15 minutes, between January $12^\text{th}$ 2018, at blockheight 503.816, to July $17^\text{th}$ 2019, at blockheight 585.844\cite{lngossip}: these snapshots were, then, aggregated into \emph{timespans}, each timespan representing a constant state of a channel from its start to its end; for the present analysis, we considered the \emph{daily snapshot} representation of the BLN, including all channels that were found to be active during that day. Each weight defining the BLN representation considered in the present work, thus, indicates the total capacity of the channels the two involved nodes have established on a given temporal snapshot - the \emph{channel capacity} coinciding with the total amount of bitcoins needed to keep it open as well as the maximum amount of tokens allowed to be exchanged across it (see fig. \ref{fig0}).

Importantly, here we do not rest upon estimates of the number of daily blocks - obtainable by considering that the time between the appearance of two subsequent blocks, in the blockchain, is Poisson distributed with an expected value of 10 minutes - but on the exact time our channels have been opened: since every channel consists of an unspent transaction output on the blockchain, we can determine the size of a channel and its opening and closing time within minutes.

\section{Methods}

\paragraph{Notation.} On a generic, daily snapshot $t$, the BLN can be described as a weighted, undirected network with total number of nodes $N^{(t)}$ and represented by an $N^{(t)}\times N^{(t)}$ symmetric matrix $\mathbf{W}^{(t)}$ whose generic entry $w_{ij}^{(t)}$ indicates the total capacity of the channels nodes $i$ and $j$ have established during the snapshot $t$ \cite{newman2003structure,newman2006structure}. Consistently, the generic entry of the BLN binary adjacency matrix $\mathbf{A}^{(t)}$ reads $a_{ij}^{(t)}=1$ if $w_{ij}^{(t)}>0$ and $a_{ij}^{(t)}=0$ otherwise: the presence of a link between any two nodes $i$ and $j$, i.e. $a_{ij}^{(t)}=1$, indicates that one or more payment channels have been opened, between the same nodes, during the snapshot $t$. As a last remark, we will focus on the largest connected component (LCC) of the BLN, throughout its entire history - the percentage of nodes belonging to it being steadily above $90\%$.

\begin{figure}[t!]
\begin{center}
\includegraphics[width=\textwidth]{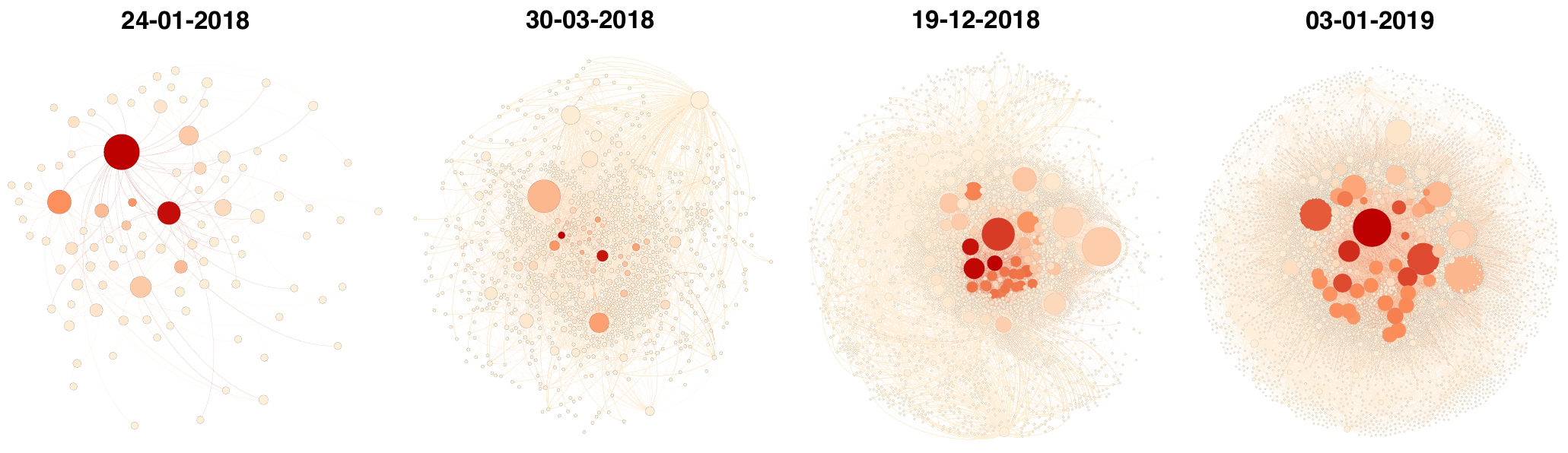}
\caption{Pictorial representation of the four snapshots of the BLN whose LCC is characterised by a number of nodes amounting at $100$, $1.000$, $3.000$, $5.000$ and corresponding to the days 24-01-2018, 30-03-2018, 19-12-2018 and 01-03-2019, respectively. The size of each node is proportional to its degree (i.e. the bigger the node, the larger its degree) while the color of each node is proportional to its strength (i.e. the darker the node, the larger its strength).}
\label{fig1}
\end{center}
\end{figure}

For the sake of illustration, we will plot our results for four snapshots, i.e. the ones whose LCC is characterised by a number of nodes amounting at $100$, $1.000$, $3.000$, $5.000$ and corresponding to the days 24-01-2018, 30-03-2018, 19-12-2018 and 01-03-2019, respectively (see fig. \ref{fig1}).

\paragraph{Degree and strength distributions.} The total number of channels (i.e. \emph{links}) that have been opened during the snapshot $t$ is provided by $L^{(t)}=\sum_{i=1}^{N^{(t)}}\sum_{j=i+1}^{N^{(t)}}a_{ij}^{(t)}$; on the other hand, the total number of channels node $i$ participates in coincides with its \emph{degree}, i.e. $k_i^{(t)}=\sum_{j(\neq i)=1}^{N^{(t)}}a_{ij}^{(t)}$. The weighted counterparts of the notions above coincide with the total weight of the network, i.e. $W^{(t)}=\sum_{i=1}^{N^{(t)}}\sum_{j=i+1}^{N^{(t)}}w_{ij}^{(t)}$, and with the total amount of money exchanged by node $i$, i.e. $s_{i}^{(t)}=\sum_{j(\neq i)=1}^{N^{(t)}}{w_{ij}^{(t)}}$, a quantity often referred to as \emph{node strength} or \emph{node capacity}.

While inspecting the functional form of the degree and strength distributions may reveal the presence of hubs, i.e. `large', single nodes, when dealing with cryptocurrencies it is of interest making a step further and inspecting the presence of `large subgraphs' of nodes. The meaning of this sentence can be made more precise upon considering the metric designed by Srinivasan et al. \cite{srinivasan2017quantifying} to measure the number of addresses required (to collude) for gathering over the $51\%$ of the overall mining power and named \emph{Nakamoto index}: a high Nakamoto coefficient indicates that many miners, or mining pools, need to combine their power to reach the $51\%$ threshold needed to take over the blockchain. Here, we adapt it to quantify a `topological' kind of majority, by defining

\begin{equation}
N_k=\min\{i\in[1\dots N]:\sum_i^Nf_i\geq 0.51\}
\end{equation}
where $f_i=k_i/2L$ and

\begin{equation}
N_s=\min\{i\in[1\dots N]:\sum_i^Nf_i\geq 0.51\}
\end{equation}
where $f_i=s_i/2W$: the first variant of the Nakamoto index can be calculated by starting from the (node with) largest degree and add them up until the condition above is satisfied; analogously, for the second variant.

\paragraph{Assortativity and hierarchy.} In order to gain insight into the higher-order structure of the BLN, we can consider the quantities known as \emph{average nearest neighbors degree}, defined as

\begin{equation}\label{equation1}
\text{ANND}_i=\frac{\sum_{j(\neq i)=1}^Na_{ij}k_j}{k_i},\quad\forall\:i
\end{equation}
and \emph{average nearest neighbors strength}, defined as

\begin{equation}\label{equation2}
\text{ANNS}_i=\frac{\sum_{j(\neq i)=1}^Na_{ij}s_j}{k_i},\quad\forall\:i;
\end{equation}
while plotting $\text{ANND}_i$ versus $k_i$ reveals the (either positive or negative) assortative character of a network, i.e. the presence of (either positive or negative) correlations between degrees, plotting $\text{ANNS}_i$ versus $k_i$ reveals the presence of (either positive or negative) correlations between degrees and strengths. On the other hand, the `cohesiveness' of the neighborhood of each node can be inspected by calculating the \emph{binary clustering coefficient}

\begin{equation}\label{equation3}
\text{BCC}_i=\frac{\sum_{j(\neq i)=1}^N\sum_{k(\neq i,j)=1}^Na_{ij}a_{jk}a_{ki}}{k_i(k_i-1)},\quad\forall\:i
\end{equation}
defined as the percentage of triangles established by any two neighbors of each node and the node itself. Its weighted counterpart reads

\begin{equation}\label{equation4}
\text{WCC}_i=\frac{\sum_{j(\neq i)=1}^N\sum_{k(\neq i,j)=1}^Nw_{ij}w_{jk}w_{ki}}{k_i(k_i-1)},\quad\forall\:i
\end{equation}
and is intended to assign a `weight' to each triangle counted by the BCC, by weighing the connections shaping it. Plotting $\text{BCC}_i$ versus $k_i$ reveals the (possibly) hierarchical character of a network, i.e. its organisation in sub-modules; plotting $\text{WCC}_i$ versus $k_i$, instead, provides a hint about the magnitude of the nodes inter-connections as a function of the nodes connectivity.

\paragraph{Disparity.} The \emph{disparity index} is defined as

\begin{equation}
Y_i=\sum_{j(\neq i)=1}^N\left[\frac{w_{ij}}{s_i}\right]^2=\frac{\sum_{j(\neq i)=1}^Nw_{ij}^2}{s_i^2}=\frac{\sum_{j(\neq i)=1}^Nw_{ij}^2}{\left[\sum_{j(\neq i)=1}^Nw_{ij}\right]^2},\quad\forall\:i
\end{equation}
and quantifies the (un)evenness of the distribution of the weights `constituting' the $i$-th strength over the $k_i$ links characterising the connectivity of node $i$. More specifically, the disparity index of node $i$ reads $Y_i=1/k_i$ in case weights are equally distributed among the connections established by it, i.e. $w_{ij}=a_{ij} s_i/k_i$, $\forall\:j$, any larger value signalling an excess concentration of weight in one or more links.

\paragraph{Centrality.} Any index measuring the centrality of a node aims at quantifying its importance in the network, according to some specific topological criterion \cite{bonacich1987power,borgatti2005centrality,newman2018networks,rodrigues2019network}. While the efforts of researchers have mainly focused on the definition of binary centrality measures, relatively little work has been done on their weighted counterparts. In what follows, we will consider possible extensions of the centrality measures employed in \cite{lin2020lightning}, i.e. the \emph{degree}, \emph{closeness}, \emph{betweenness} and \emph{eigenvector} centrality:

\begin{itemize}
\item the degree centrality \cite{newman2018networks,rodrigues2019network} of node $i$ coincides with its degree, normalized by the maximum attainable value, i.e. $\text{DC}_i=k_i/(N-1)$: the strength centrality of node $i$ generalises it by simply replacing the total number of `node-specific' connections with the total `node-specific' weight. In what follows we will consider the (simpler) definition

\begin{equation}
\text{WDC}_i=s_i,\quad\forall\:i
\end{equation}
from which it follows that the most central node, according to the strength variant, is the one characterised by the largest percentage of weight `embodied' by (the totality of) its connections;

\item the closeness centrality \cite{newman2018networks,rodrigues2019network} of node $i$ is defined as $\text{CC}_i=(N-1)/\sum_{j(\neq i)=1}^Nd_{ij}$ where $d_{ij}$ is the topological distance between nodes $i$ and $j$, i.e. the length of any shortest path connecting them. The definition of weighted closeness centrality of node $i$ is based on the redefinition of shortest path length which, in turn, rests upon the redefinition of \emph{weighted distance} between any two nodes, i.e. $d_{ij}^{(w)}$. Possible variants of the latter one read $d_{ij}^{(w)}=\min\{w_{ih}+\dots+w_{hj}\}$ and $d^{(w)}_{ij}=\min\left\{\frac{1}{w_{ih}}+\dots+\frac{1}{w_{hj}}\right\}$ where $h$ indexes the intermediary vertices lying on the path between $i$ and $j$, $w_{ih}\dots w_{hj}$ are the weights of the corresponding edges and the extremum is taken over all paths between $i$ and $j$. Naturally, the meaning changes along with the chosen definition: while the first one describes any two nodes as `closer', the smaller the weights of the intermediate connections, the opposite is true when the second one is considered. Hereby, we opt for the following definition of weighted closeness centrality

\begin{equation}
\text{WCC}_i=\frac{N-1}{\sum_{j(\neq i)=1}^Nd^{(w)}_{ij}},\quad\forall\:i
\end{equation}
with $d^{(w)}_{ij}=\min\left\{\frac{1}{w_{ih}}+\dots+\frac{1}{w_{hj}}\right\}$. This choice also implies that once the path connecting nodes $i$ and $j$ has been individuated, the WCC is nothing else that the harmonic mean of the weights constituting it;

\begin{figure}[t!]
\begin{center}
\includegraphics[width=\textwidth]{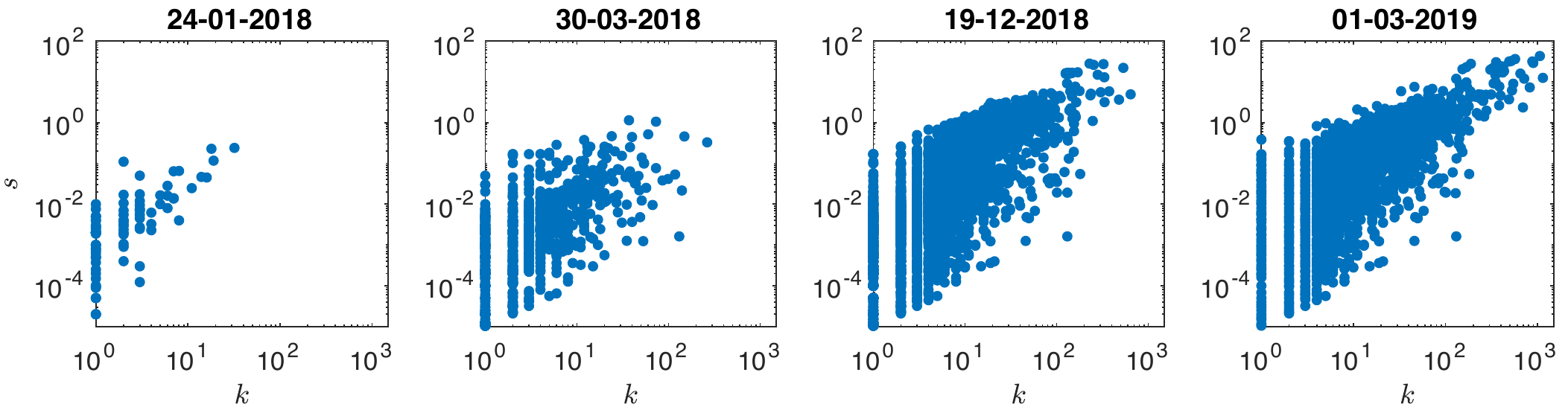}
\caption{Scattering the strength sequence versus the degree sequence reveals the presence of positive correlations between the two sets of quantities: the Pearson coefficient describing them, on our usual four snapshots, amounts at $r=0.84,0.42,0.66,0.80$, respectively.}
\label{fig2}
\end{center}
\end{figure}

\item the betweenness centrality \cite{newman2018networks,newman2005measure,pfitzner2013betweenness,bonacich2007some} of node $i$ is given by $\text{BC}_{i}=\sum_{s(\neq i)=1}^N\sum_{t(\neq i,s)=1}^N\sigma_{st}(i)/\sigma_{st}$ where $\sigma_{st}$ is the total number of shortest paths between node $s$ and $t$ and $\sigma_{st}(i)$ is the number of shortest paths between nodes $s$ and $t$ that pass through node $i$. The weighted counterpart of it can be defined as

\begin{equation}
\text{WBC}_{i}=\sum_{s(\neq i)=1}^N\sum_{t(\neq i,s)=1}^N\frac{\sigma_{st}^{(w)}(i)}{\sigma_{st}^{(w)}},\quad\forall\:i
\end{equation}
where, now, $\sigma_{st}^{(w)}$ is the total number of weighted shortest paths between nodes $s$ and $t$ and $\sigma_{st}^{(w)}(i)$ is the number of weighted shortest paths between nodes $s$ and $t$ that pass through node $i$;

\item the eigenvector centrality \cite{newman2018networks,liu2016locating,bonacich2007some} of node $i$ is defined as the $i$-th element of the eigenvector corresponding to the largest eigenvalue of the binary adjacency matrix - whose existence is guaranteed in case the Perron-Frobenius theorem holds true. According to the definition above, a node with large eigenvector centrality is connected to other `well connected' nodes. Such a definition can be extended by considering the $\text{WEC}_i$, defined as the $i$-th element of the eigenvector corresponding to the largest eigenvalue of the weighted adjacency matrix.
\end{itemize}

The Gini coefficient has been introduced to quantify `inequality' in wealth distribution \cite{morgan1962anatomy,crucitti2006centrality} and ranges between 0 and 1, a larger Gini coefficient indicating a larger `(un)evenness' of the income distribution:

\begin{equation}
G_c=\frac{\sum_{i=1}^N\sum_{j=1}^N|c_i-c_j|}{2N\sum_{i=1}^N{c_i}};
\end{equation}
hereby, we apply it to the several definitions of centrality provided above. As a general comment, we would like to stress that a non-normalized centrality measure cannot be employed to compare nodes, across different configurations, in a fully consistent way. However, if our only interest is that of quantifying the (un)evenness of the distribution of our centrality measures, the absence of a normalization term does not make any difference: in fact, the Gini coefficient is not affected by it.

\begin{figure}[t!]
\begin{center}
\includegraphics[width=\textwidth]{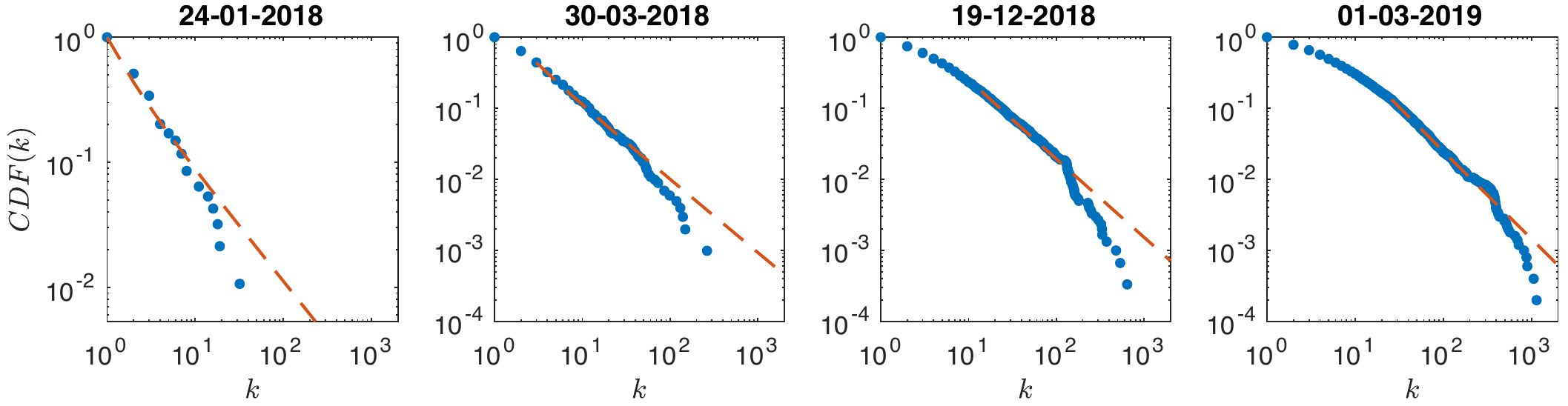}
\caption{Cumulative density function of the degrees, for our usual four snapshots. The support of the distribution has become broader as the BLN has evolved. Fitting a power-law $\text{PDF}(k)=(\alpha-1)k_{min}^{\alpha-1}k^{-\alpha}$ on the data, by running the code released by Clauset et al. \cite{clauset2009power}, returns values amounting at $\alpha=1.9,2.0,2.1,2.2$ and $k_{min}=1,3,14,26$ (see the dashed lines) while the Kolmogorov-Smirnov test returns the p-values $p=0.02,0.03,0.04,0.5$. Hence, the null hypothesis that the degrees are distributed according to a power-law is never rejected, at the $1\%$ significance level - while it is, for the first three snapshots, at the $5\%$ significance level. Overall, the null hypothesis that the degrees are distributed according to a power-law is not rejected for the $85\%$ of the total number of snapshots, at the $1\%$ significance level, and for the $71\%$ of the total number of snapshots, at the $5\%$ significance level.}
\label{fig3}
\end{center}
\end{figure}

\paragraph{Small-world -ness.} The study of the BLN centralisation can be approached from a slightly different perspective by asking if the BLN is (increasingly) becoming a small-world system \cite{watts1998collective,latora2001efficient,amara2011classes}. The usual way of proceeding to answer such a question prescribes to check if

\begin{equation}
\overline{d}=\frac{\sum_{i=1}^N\sum_{j(\neq i)=1}^Nd_{ij}}{N(N-1)}\sim\ln N
\end{equation}
i.e. if the average path length grows logarithmically with the number of nodes and if the average clustering coefficient $\overline{\text{BCC}}_i=\sum_{i=1}^N\text{BCC}_i/N$ is larger than the one predicted by an \emph{Undirected Random Graph Model} (URGM) tuned to reproduce the empirical density of links. Recently, however, it has been argued that the same question can be answered by considering the quantity named \emph{global efficiency}, defined as

\begin{equation}
E_g=\frac{\sum_{i=1}^N\sum_{j(\neq i)=1}^Nd^{-1}_{ij}}{N(N-1)},
\end{equation}
understood as an indicator of the `traffic capacity' of a network and, quite remarkably, not affected by the analytical problems suffered by the average path length \cite{latora2001efficient} - potentially diverging due to the presence of couples of nodes belonging to disconnected components. Latora et al. \cite{latora2001efficient} have also defined a \emph{local efficiency} as

\begin{equation}
E_l=\frac{1}{N}\sum_{i=1}^NE(\mathbf{G}_i),
\end{equation}
a quantity that can be evaluated by, first, calculating the efficiency of the subgraph induced by the nearest neighbors of each node, upon removing it and, then, averaging such numbers. Latora et al. \cite{latora2001efficient} have argued that while $E_g$ plays a role analogous to the inverse of the average path length, $E_l$ plays a role analogous to the average clustering coefficient: hence, small-world networks should have both a large $E_g$ and a large $E_l$, i.e. should be very efficient in allowing nodes to communicate in both a global and a local fashion.

\paragraph{Core-periphery detection.} As it has emerged quite clearly from the binary analysis of the BLN, just inspecting the evolution of centrality measures can return a too simplistic picture of the network under consideration. For this reason, we have checked for the presence of mesoscopic `centralised' structures such as the \emph{core-periphery} one, composed by a densely-connected subgraph of nodes surrounded by a periphery of loosely-connected vertices. In order to do so, we have implemented the approach recently proposed in \cite{marchese2022detecting} and prescribing to minimize the score function

\begin{equation}
\mathscr{W}_\sslash=\sum_{w_\bullet\geq w_\bullet^*}\sum_{w_\circ\geq w_\circ^*}\frac{\big(\binom{V_\bullet}{w_\bullet}\big)\big(\binom{V_\circ}{w_\circ}\big)\big(\binom{V-V_\bullet-V_\circ}{W-w_\bullet-w_\circ}\big)}{\big(\binom{V}{W}\big)}
\end{equation}
known as \emph{bimodular surprise}\footnote{The Python package for surprise optimization, called `SurpriseMeMore', is freely downloadable at the following URL: https://github.com/EmilianoMarchese/SurpriseMeMore.}; here, $V=N(N-1)/2$ is the total number of node pairs, $W=\sum_{i=1}^N\sum_{j=i+1}^{N}w_{ij}$ is the total weight of the network, $V_\bullet$ is the number of node pairs in the core portion of the network, $V_\circ$ is the number of node pairs in the periphery portion of the network, $w_\bullet^*$ is the observed number of core links and $w_\circ^*$ is the observed number of periphery links. From a technical point of view, $\mathscr{W}_\sslash$ is the p-value of a multivariate negative hypergeometric distribution and the multiset notation, according to which $\big(\binom{V_\bullet}{w_\bullet}\big)=\binom{V_\bullet+w_\bullet-1}{w_\bullet}$ allows $\mathscr{W}_\sslash$ to be compactly rewritten in a way that nicely mirrors that of its binary counterpart \cite{dejeude2019detecting,marchese2022detecting}.

\begin{figure}[t!]
\begin{center}
\includegraphics[width=\textwidth]{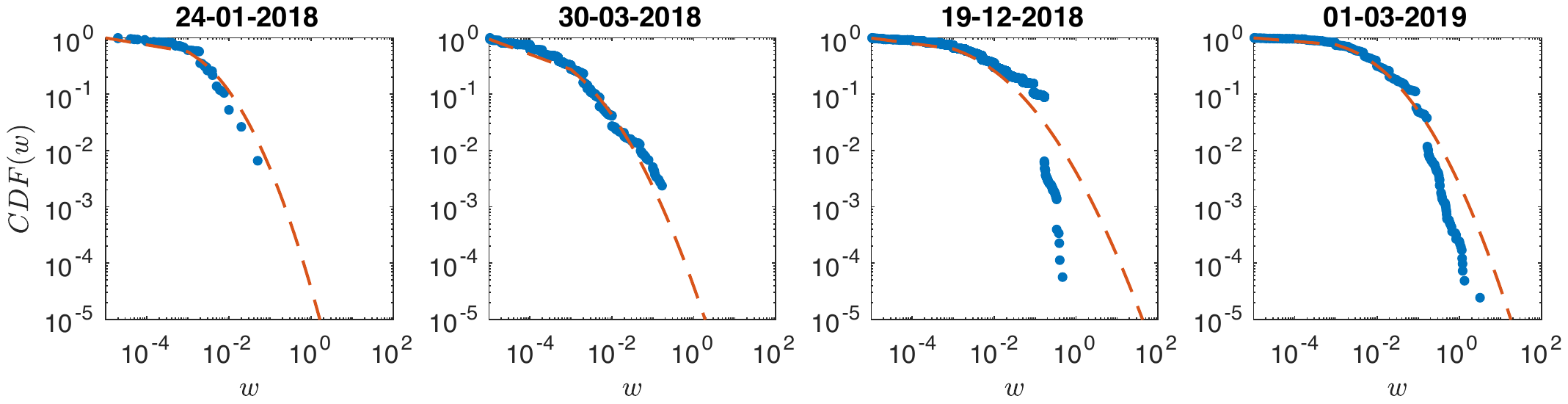}
\caption{Cumulative density function of the weights, for our usual four snapshots. The support of the distribution has become slightly broader as the BLN has evolved. Fitting a log-normal distribution $\text{PDF}(w)=(w\sigma\sqrt{2\pi})^{-1}e^{-\frac{(\ln w-\mu)^2}{2\sigma^2}}$ on the data reveals that, at both the $1\%$ and the $5\%$ significance levels, the Kolmogorov-Smirnov test does not reject the hypothesis that weights are log-normally distributed when $N<94$ (i.e. from the fourth day to the twelfth day); for our four snapshots, instead, the hypothesis is rejected - notice that day 24-01-2018 is the thirteenth. Each dashed line represents the best fit for the specific snapshot: the corresponding parameters read $\mu=-6.6,-8.1,-6.1,-5.5$ and $\sigma=1.7,2.1,2.3,2.0$.}
\label{fig4}
\end{center}
\end{figure}

\paragraph{Benchmarking the observations.} Along the guidelines of the analysis carried out in \cite{lin2020lightning}, in what follows we benchmark our observations by employing the recently-proposed null model called $\text{CReM}_\text{A}$ - the acronym standing for \emph{Conditional Reconstruction Model A} \cite{parisi2020faster,vallarano2021fast} - that allows binary and weighted constraints to be defined in a disentangled fashion. From a purely theoretical point of view, it is defined by the maximisation of the \emph{conditional Shannon entropy}

\begin{equation}
S(\mathscr{W}|\mathscr{A})=-\sum_{\mathbf{A}\in\mathbb{A  }}P(\mathbf{A})\int_{\mathbb{W}_\mathbf{A}}Q(\mathbf{W}|\mathbf{A})\ln Q(\mathbf{W}|\mathbf{A})d\mathbf{W}
\end{equation}
constrained to reproduce the strengths $\{s_i\}_{i=1}^N$; the (conditional) weighted distribution output by such an optimization procedure reads

\begin{equation}
Q(\mathbf{W}|\mathbf{A})=\frac{e^{-H(\mathbf{W})}}{Z_\mathbf{A}}=\prod_{i=1}^N\prod_{j=i+1}^Nq_{ij}(w_{ij}|a_{ij})=\prod_{i=1}^N\prod_{j=i+1}^N(\beta_i+\beta_j)^{a_{ij}}e^{-(\beta_i+\beta_j)w_{ij}};
\end{equation}
notice the conditional character of the distribution above, embodied by the term $a_{ij}$ at the exponent - as a simple consistency check, the probability that $w_{ij}=0$ in case there is no link is $q(w_{ij}=0|a_{ij}=0)=1$ as it should be. The vector of parameters $\{\beta_i\}_{i=1}^N$ defining the distribution above can be estimated via a (generalized) \emph{likelihood maximisation} procedure \cite{parisi2020faster} that leads to the system of $N$ equations

\begin{equation}\label{crema}
s_i=\sum_{j(\neq i)=1}^N\langle w_{ij}\rangle=\sum_{j(\neq i)=1}^N\frac{p_{ij}}{\beta_i+\beta_j},\quad\forall\:i;
\end{equation}
the coefficients $\{p_{ij}\}_{i,j=1}^N$, instead, are treated as `prior information' and, as such, left  `untouched' by the estimation procedure above. In a sense, we are free to combine the (conditional) weighted distribution above with the purely binary probability mass function `best' encoding the available information about the network structure. In what follows, we have considered

\begin{figure}[t!]
\begin{center}
\includegraphics[width=\textwidth]{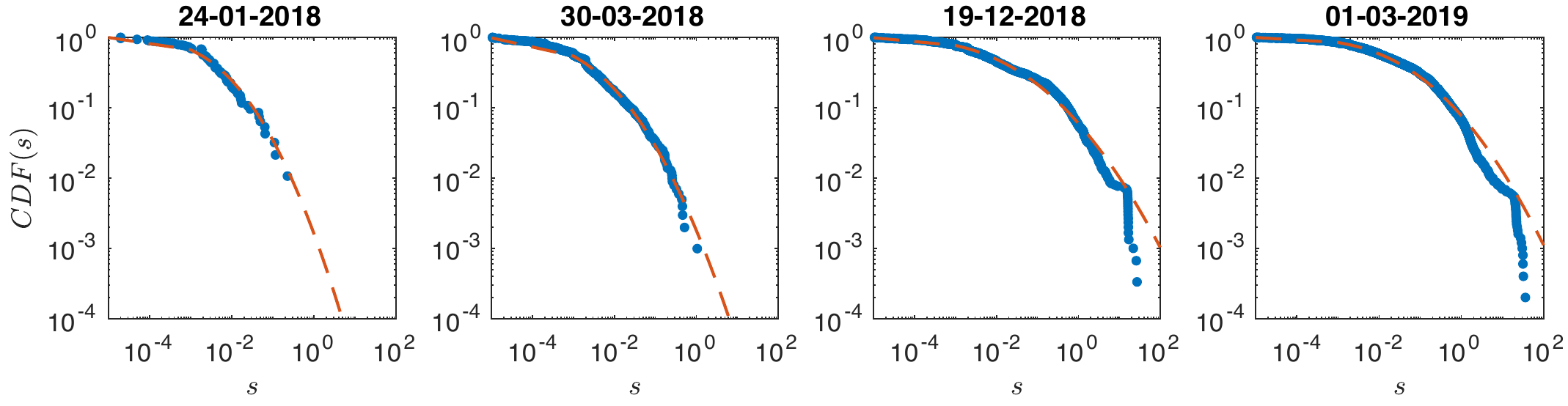}
\caption{Cumulative density function of the strengths, for our usual four snapshots. The support of the distribution is enlarged of a few orders of magnitude during the BLN history. A log-normal distribution $\text{PDF}(s)=(s\sigma\sqrt{2\pi})^{-1}e^{-\frac{(\ln s-\mu)^2}{2\sigma^2}}$, fitted on the data, lets the Kolmogorov-Smirnov test returns the p-values $p=0.061,0.006, 4.4\cdot 10^{-7}$ and $1.6\cdot 10^{-7}$. Hence, the null hypothesis that strengths are log-normally distributed is not rejected for the first snapshot while it is for the last three ones - at both significance levels. Overall, the null hypothesis that the strengths are distributed according to a log-normal is not rejected for the $16\%$ of the total number of snapshots, at the $1\%$ significance level, and for the $5\%$ of the total number of snapshots, at the $5\%$ significance level. Each dashed line represents the best fit for the specific snapshot: the corresponding parameters read $\mu=-6.0,-6.7,-4.7,-4.0$ and $\sigma=2.0,2.3,3.0,2.8$.}
\label{fig5}
\end{center}
\end{figure}

\begin{itemize}
\item the one defining the \emph{Undirected Binary Configuration Model} (UBCM) and following from the maximisation of the traditional Shannon entropy $S=-\sum_\mathbf{A}P(\mathbf{A})\ln P(\mathbf{A})$ constrained to reproduce the degrees $\{k_i\}_{i=1}^N$: the UBCM captures the idea that the probability for any two nodes to establish a connection (solely) depends on their degrees and can be fully determined by solving the $N$ equations

\begin{equation}\label{ubcm}
k_i=\sum_{j(\neq i)=1}^Np_{ij}^\text{UBCM}=\sum_{j(\neq i)=1}^N\frac{x_ix_j}{1+x_ix_j},\quad\forall\:i;
\end{equation}

\item the deterministic recipe $p_{ij}\equiv a_{ij}$, $\forall\:i<j$, accounting for the case in which the prior knowledge concerns the entire network topological structure, now treated as given.
\end{itemize}

While, in the first case, the generic set of coefficients $\{p_{ij}\}_{i,j=1}^N$ is instantiated upon identifying $p_{ij}\equiv p_{ij}^\text{UBCM}$, $\forall\:i<j$, in the second one the, identification simply reads $p_{ij}\equiv a_{ij}$, $\forall\:i<j$; in both cases, the resolution of the related system of equations, carried out via the Python package called `NEMTROPY'\footnote{The acronym stands for `Network Entropy Maximization: a Toolbox Running On Python' and the package is freely downloadable at the following URL: https://pypi.org/project/NEMtropy/.}, leads us to numerically determine the corresponding vector of parameters $\{\beta_i\}_{i=1}^N$.

Benchmarking a set of observations ultimately boils down at verifying their `compatibility' with the predictions output by a chosen null model, by testing their statistical significance against the null model itself. To this aim, one can proceed as follows: first, sampling the ensemble induced by the chosen null model, by generating a `sufficiently large' number of configurations (in all our experiments, 100); second, calculating the value of any quantity of interest over each configuration; third, deriving the corresponding ensemble CDF. At this point, a p-value remains naturally defined; in what follows, we will employ it to carry out one-tailed tests. Whenever tests of this kind are considered, one may be interested in calculating either the (ensemble) probability $Q(X\geq X^*)$ of observing a value, for the quantity of interest $X$, that is \emph{larger} than the empirical one, $X^*$, or the (ensemble) probability $Q(X\leq X^*)$ of observing a value, for the quantity of interest $X$, that is \emph{smaller} than it; in both cases, if such a probability is found to be smaller than a given threshold, the quantity is deemed as statistically significant, hence incompatible with the description provided by the chosen null model - which (significantly) underestimates or overestimates it, respectively.

\begin{figure}[t!]
\begin{center}
\includegraphics[width=0.5\textwidth]{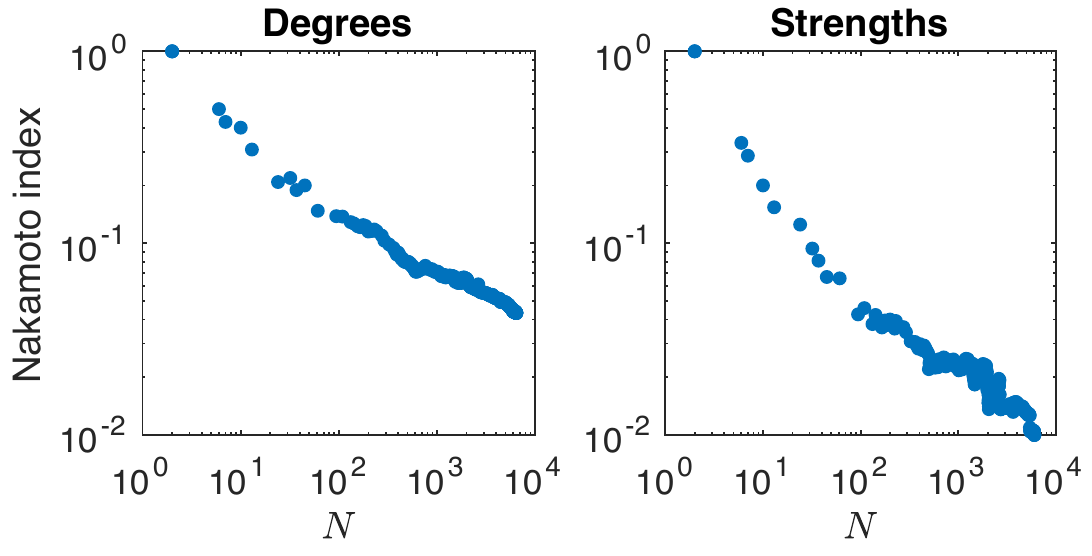}
\caption{Evolution of the Nakamoto index for the degrees and the strengths, plotted versus the total number of nodes: as the size of the system enlarges, the percentage of nodes `providing' the $51\%$ of the total number of links/the total weight progressively reduces, an evidence pointing out that nodes embodying a `topological' kind of majority indeed appear. Moreover, the total weight seems to be distributed less evenly than the total number of connections.}
\label{fig6}
\end{center}
\end{figure}

\section{Results}

\paragraph{Degree and strength distributions.} Giving a look at the four snapshots depicted in fig. \ref{fig1} reveals the presence of a large heterogeneity, with nodes having a large degree/strength co-existing with nodes having a small degree/strength; moreover, while nodes with a large degree also have a large strength (i.e. larger nodes are also darker), small, dark nodes can be observed as well: in other words, an overall positive correlation between degrees and strengths co-exists with a large variability of the strength values - especially for what concerns the nodes with a small connectivity (see fig. \ref{fig2}).

As a first empirical analysis, we have inspected the functional form of the degree distribution for four distinct snapshots, i.e. the days 24-01-2018, 30-03-2018, 19-12-2018 and 01-03-2019; to this aim, we have plotted the cumulative density function (CDF), defined as $\text{CDF}(k)=\sum_{h\geq k}f(h)$ where $f(h)$ is the fraction of nodes whose degree is $h$. As shown in fig. \ref{fig3}, the degree distribution becomes broader as the BLN evolves; moreover, running the code released by Clauset et al. \cite{clauset2009power} to fit the functional form $\text{PDF}(k)=(\alpha-1)k_{min}^{\alpha-1}k^{-\alpha}$ on the data returns the values $\alpha=1.9,2.0,2.1,2.2$ and $k_{min}=1,3,14,26$ while the Kolmogorov-Smirnov test returns the p-values $p=0.02,0.03,0.04,0.5$. Hence, the null hypothesis that the degrees are distributed according to a power-law is never rejected, at the $1\%$ significance level - while it is, for the first three snapshots, at the $5\%$ significance level. Overall, the null hypothesis that the degrees are distributed according to a power law is not rejected for the $85\%$ of the total number of snapshots, at the $1\%$ significance level, and for the $71\%$ of the total number of snapshots, at the $5\%$ significance level.

As a second empirical analysis, we have calculated the evolution of the CDF of the weights, defined as $\text{CDF}(w)=\sum_{v\geq w}f(v)$. Analogously to what has been observed for the degrees, even the support of the weight distribution has broadened throughout the entire BLN history (see fig. \ref{fig4}), although to a lesser extent. Fitting a log-normal distribution, whose functional form reads $\text{PDF}(w)=(w\sigma\sqrt{2\pi})^{-1}e^{-\frac{(\ln w-\mu)^2}{2\sigma^2}}$, on the data reveals that, at both the $1\%$ and the $5\%$ significance levels, the Kolmogorov-Smirnov test does not reject the hypothesis that weights are log-normally distributed when $N<94$ (i.e. from the fourth day to the twelfth day). For our four snapshots, instead, the hypothesis is rejected - notice that day 24-01-2018 is the thirteenth.

As a third empirical analysis, we have considered the evolution of the CDF of the strengths, defined as $\text{CDF}(s)=\sum_{t\geq s} f(s)$. The support of the distribution is enlarged of a few orders of magnitude during the BLN history (see fig. \ref{fig5}). Analogously to the case of the weights, we have fitted a log-normal distribution, whose functional form reads $\text{PDF}(s)=(s\sigma\sqrt{2\pi})^{-1}e^{-\frac{(\ln s-\mu)^2}{2\sigma^2}}$, on the data: while the the Kolmogorov-Smirnov test returns the p-values $p=0.061,0.006,4.4\cdot 10^{-7}$ and $1.6\cdot 10^{-7}$, hence does not reject the hypothesis that strengths are log-normally distributed, at both significance levels, for the first snapshot considered here, it does so for the other three ones - an evidence seemingly indicating that, quite early in its history, the BLN has started deviating more and more from the picture provided by the distribution tested here. Overall, the null hypothesis that the strengths are distributed according to a log-normal is not rejected for the $16\%$ of the total number of snapshots, at the $1\%$ significance level, and for the $5\%$ of the total number of snapshots, at the $5\%$ significance level.

The picture provided by the three distributions above can be complemented by the information provided by the Nakamoto index (see fig. \ref{fig6}). As its evolution clearly shows, the percentage of nodes `providing' the $51\%$ of the total number of links/the total weight progressively reduces, as the BLN size enlarges: in particular, the total weight seems to be distributed less evenly than the total number of connections - as fewer nodes are needed to embody the (same) required percentage. This seems to confirm the appearance of nodes constituting the aforementioned `topological' majority.

\begin{figure}[t!]
\begin{center}
\includegraphics[width=\textwidth]{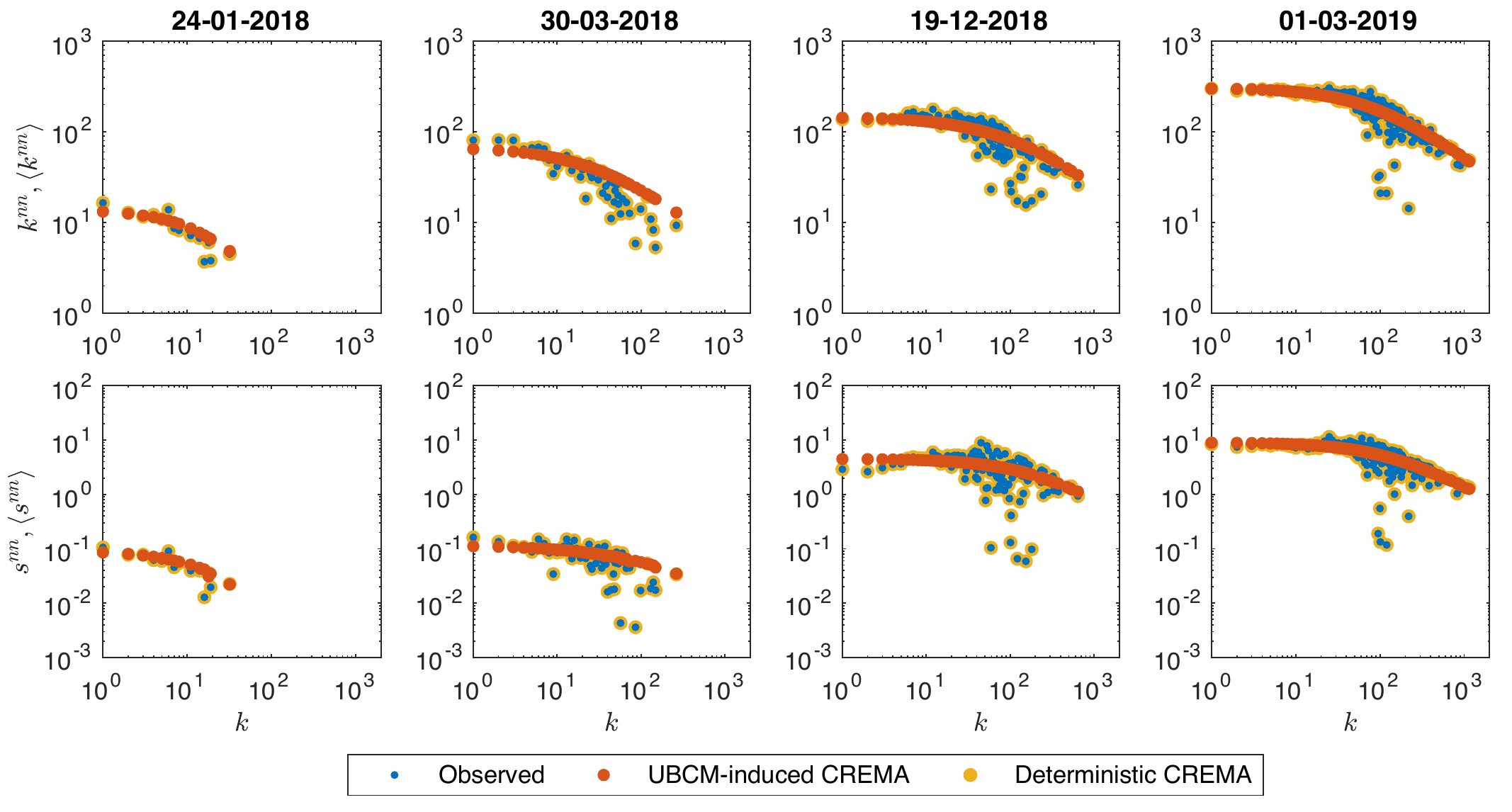}
\caption{$\text{ANND}_i$, $\langle\text{ANND}_i\rangle$ values scattered versus $k_i$ (upper panels) and $\text{ANNS}_i$, $\langle\text{ANNS}_i\rangle$ values scattered versus $k_i$ (bottom panels) for our usual four snapshots (all trends are averaged over the classes of nodes with the same degree). Both trends clearly signal a disassortative behavior, i.e. nodes with a large degree are (preferentially) connected to nodes with a small degree/small strength and viceversa. While the UBCM-induced $\text{CReM}_\text{A}$ model successfully captures such a disassortative trend, the deterministic $\text{CReM}_\text{A}$ model reproduces both the ANND and the ANNS values exactly.}
\label{fig7}
\end{center}
\end{figure}

\paragraph{Assortativity and hierarchy.} Plotting the values of the average nearest neighbors degree versus the degrees reveals the disassortative character of the BLN, i.e. the presence of negative correlations between the degrees: in other words, nodes with a large degree are (preferentially) connected to nodes with a small degree and viceversa. To be noticed that the UBCM-induced $\text{CReM}_\text{A}$ model successfully captures such a trend, indicating that the information encoded into the degree sequence, leading to

\begin{equation}
\langle\text{ANND}_i\rangle\simeq\frac{\sum_{j(\neq i)=1}^N\langle a_{ij}\rangle\langle k_j\rangle}{\langle k_i\rangle}=\frac{\sum_{j(\neq i)=1}^Np_{ij}k_j}{k_i},\quad\forall\:i
\end{equation}
with $p_{ij}\equiv p_{ij}^\text{UBCM}$, $\forall\:i<j$ and where the symbol $\simeq$ indicates that we have approximated the expected value of a ratio as the ratio of the expected values, is enough to account for the correlations between the degrees as well. An analogous decreasing trend characterises the values of the average nearest neighbors strength when plotted versus the degrees, i.e. nodes with a large degree are (preferentially) connected to nodes with a small strength and viceversa; as for its binary counterpart, the UBCM-induced $\text{CReM}_\text{A}$ model successfully reproduces the empirical ANNS values, indicating that the information encoded into the degree and the strength sequences, leading to

\begin{figure}[t!]
\begin{center}
\includegraphics[width=\textwidth]{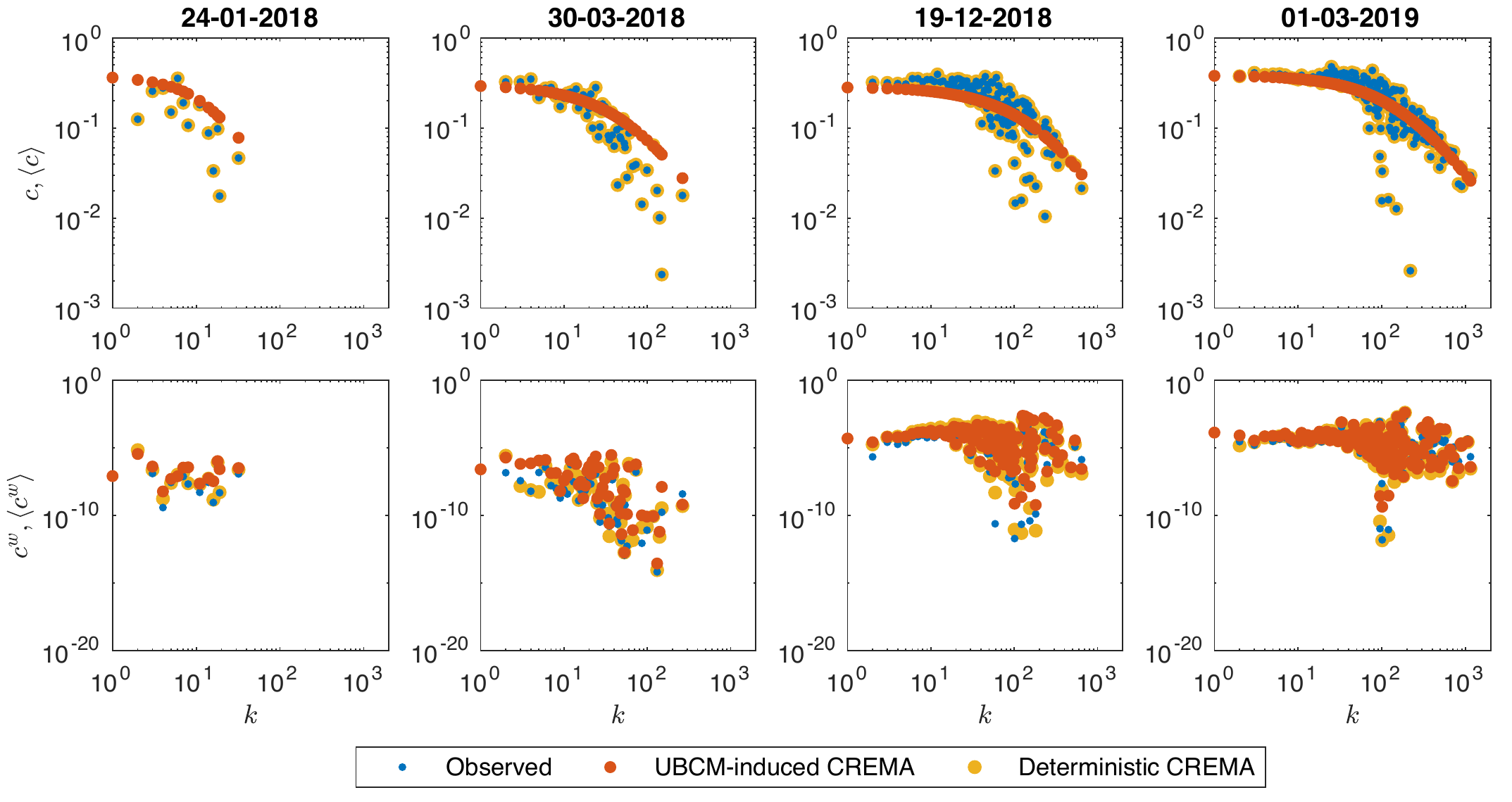}
\caption{$\text{BCC}_i$, $\langle\text{BCC}_i\rangle$ values scattered versus $k_i$ (upper panels) and $\text{WCC}_i$, $\langle\text{WCC}_i\rangle$ values scattered versus $k_i$ (bottom panels) for our usual four snapshots (all trends are averaged over the classes of nodes with the same degree). While the trend of the BCC clearly signals a hierarchical behavior, i.e. the tendency of nodes with a larger degree to participate into a smaller number of connected triples than nodes with a smaller degree and viceversa, this does not seem to be the case for the WCC values when plotted versus the degrees. While the UBCM-induced $\text{CReM}_\text{A}$ model successfully captures both trends, the deterministic $\text{CReM}_\text{A}$ model reproduces only the BCC values exactly.}
\label{fig8}
\end{center}
\end{figure}

\begin{equation}
\langle\text{ANNS}_i\rangle\simeq\frac{\sum_{j(\neq i)=1}^N\langle a_{ij}\rangle\langle s_j\rangle}{\langle k_i\rangle}=\frac{\sum_{j(\neq i)=1}^Np_{ij}s_j}{k_i},\quad\forall\:i
\end{equation}
with $p_{ij}\equiv p_{ij}^\text{UBCM}$, $\forall\:i<j$ successfully accounts for the correlations between the degrees and the strengths as well (see fig. \ref{fig7}). On the other hand, plotting the values of the clustering coefficient versus the degrees reveals the hierarchical character of the BLN: nodes with a larger degree tend to participate into a smaller number of connected triples than nodes with a smaller degree and viceversa; the UBCM-induced $\text{CReM}_\text{A}$ model, leading to

\begin{figure}[t!]
\begin{center}
\includegraphics[width=\textwidth]{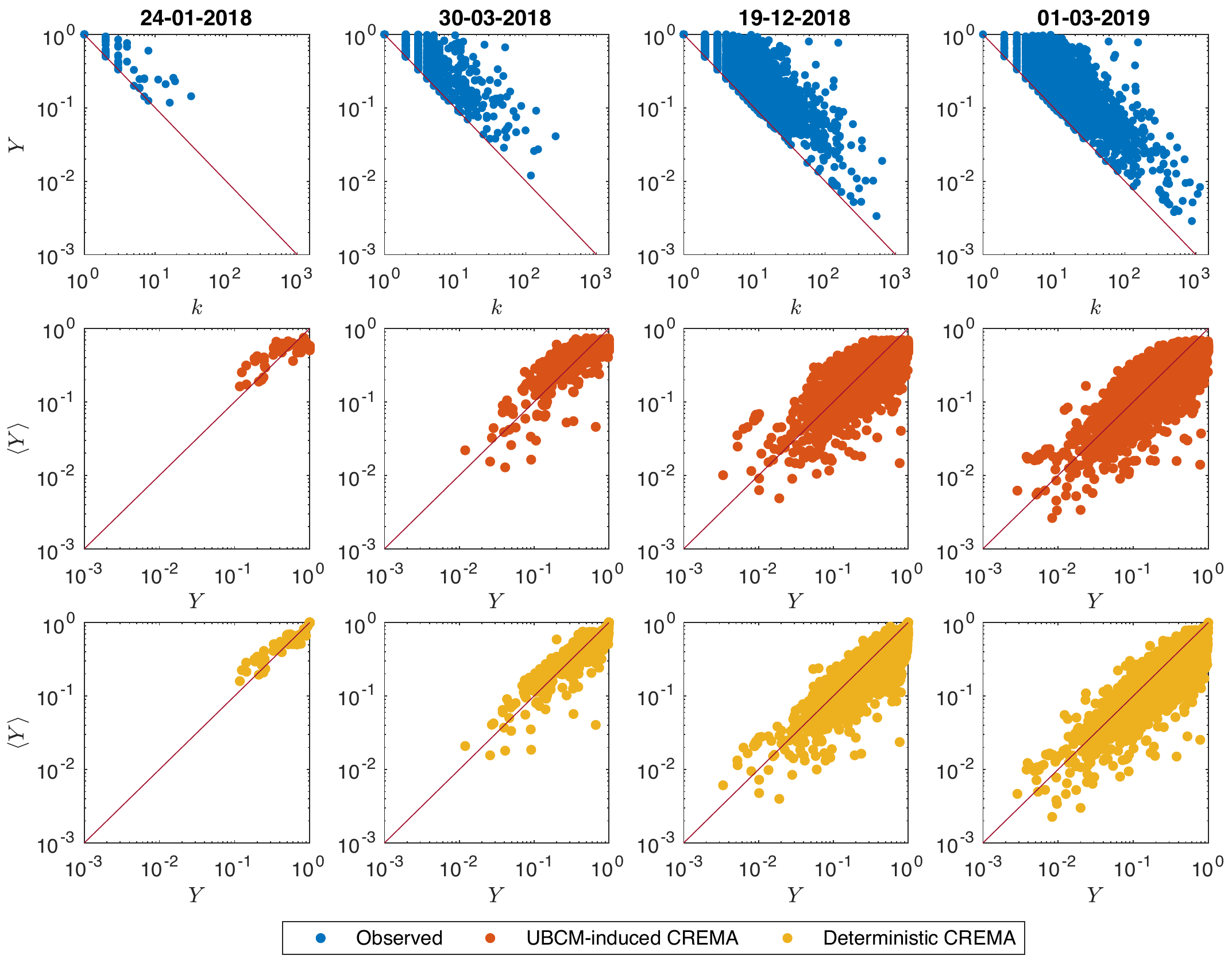}
\caption{Upper panels: empirical disparity values scattered versus the degrees, for our usual four snapshots. As the plots reveal, the vast majority of strength values is not evenly distributed across the connections characterising each node, i.e. $Y_i>1/k_i$ for the vast majority of nodes. Middle panels: expected disparity values output by the UBCM-induced $\text{CReM}_\text{A}$ model scattered versus the empirical disparity values. Bottom panels: expected disparity values output by the deterministic $\text{CReM}_\text{A}$ model scattered versus the empirical disparity values. The empirical disparity values are, generally speaking, in agreement with our benchmark models; however, the percentage of nodes for which $Q(Y_i\geq Y_i^*)<0.05$, for our usual four snapshots, amounts at $0\%,3.0\%,9.1\%,11\%$ for the UBCM-induced $\text{CReM}_\text{A}$ model and at $0\%,5.6\%,15\%,17\%$ for the deterministic $\text{CReM}_\text{A}$ model: in other words, the percentage of nodes whose empirical disparity is significantly larger than predicted by one of the two null models considered here is rising throughout the entire BLN history - i.e. its vertices increasingly `favor' some of the links surrounding them.}
\label{fig9}
\end{center}
\end{figure}

\begin{equation}
\langle\text{BCC}_i\rangle\simeq\frac{\sum_{j(\neq i)=1}^N\sum_{k(\neq i,j)=1}^N\langle a_{ij}\rangle\langle a_{jk}\rangle\langle a_{ki}\rangle}{\sum_{j(\neq i)=1}^N\sum_{k(\neq i,j)=1}^N\langle a_{ij}\rangle\langle a_{ik}\rangle}=\frac{\sum_{j(\neq i)=1}^N\sum_{k(\neq i,j)=1}^Np_{ij}p_{jk}p_{ki}}{\sum_{j(\neq i)=1}^N\sum_{k(\neq i,j)=1}^Np_{ij}p_{ik}},\quad\forall\:i
\end{equation}
with $p_{ij}\equiv p_{ij}^\text{UBCM}$, $\forall\:i<j$ is able to capture such a trend as well. The same decreasing trend, instead, does not characterise the values of the weighted clustering coefficient when plotted versus the degrees which, instead, appears as rather flat - interestingly, this is no longer true when the weighted clustering coefficient values are plotted versus the strengths: in this case, a clear rising trend is visible, signalling that nodes with a larger strength tend to participate into `heavier' connected triples of nodes. Again, the UBCM-induced $\text{CReM}_\text{A}$ model, predicting

\begin{equation}
\langle\text{WCC}_i\rangle\simeq\frac{\sum_{j(\neq i)=1}^N\sum_{k(\neq i,j)=1}^N\langle w_{ij}\rangle\langle w_{jk}\rangle\langle w_{ki}\rangle}{\sum_{j(\neq i)=1}^N\sum_{k(\neq i,j)=1}^N\langle a_{ij}\rangle\langle a_{ik}\rangle}=\frac{\sum_{j(\neq i)=1}^N\sum_{k(\neq i,j)=1}^N\langle w_{ij}\rangle\langle w_{jk}\rangle\langle w_{ki}\rangle}{\sum_{j(\neq i)=1}^N\sum_{k(\neq i,j)=1}^Np_{ij}p_{ik}},\quad\forall\:i
\end{equation}
with $p_{ij}\equiv p_{ij}^\text{UBCM}$, $\forall\:i<j$ successfully reproduces the empirical WCC values, indicating that the information encoded into the degree and the strength sequences successfully accounts for the behavior of third-order properties as well (see fig. \ref{fig8}).

\begin{figure}[t!]
\begin{center}
\includegraphics[width=\textwidth]{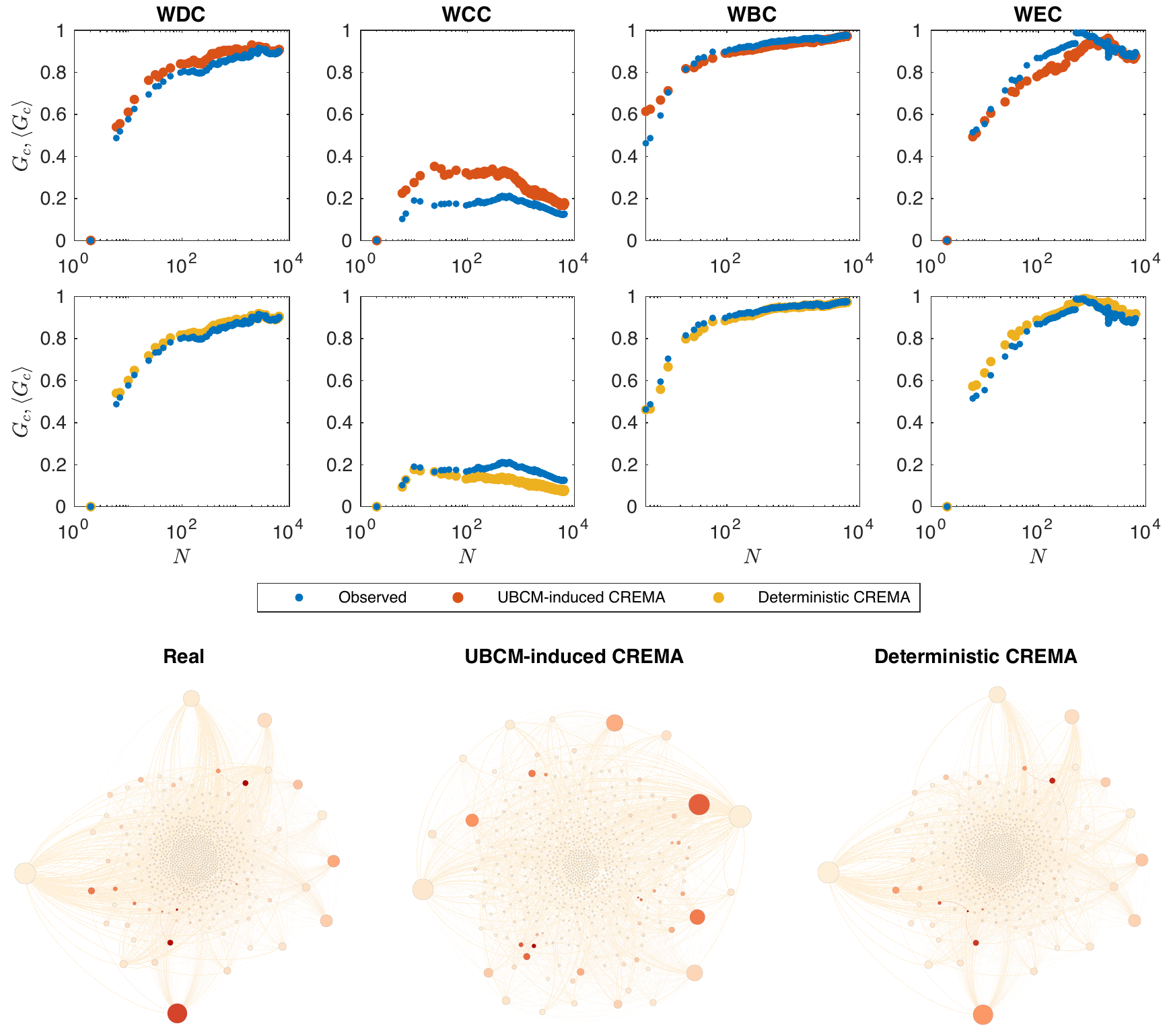}
\caption{Upper and middle panels: evolution of the Gini coefficient of the empirical weighted centrality values (blue dots) and of its expected counterpart, under the UBCM-induced $\text{CReM}_\text{A}$ model (red dots, upper panels) and the deterministic $\text{CReM}_\text{A}$ model (yellow dots, bottom panels). The rise of the Gini coefficient for the weighted degree, betweenness and eigenvector centrality points out that the distribution of centrality measures becomes increasingly uneven while the flat trend characterising the evolution of the closeness centrality confirms what has been observed in the purely binary case: the aforementioned `hubs' ensure the vast majority of nodes to be reachable (hence, to be close to each other) quite easily. Generally speaking, our null models tend to significantly underestimate both the weighted betweenness centrality and the weighted eigenvector centrality, signalling the presence of fewer-nodes-with-heavier-connections than predicted by chance. Bottom panels: comparison between the BLN on day 20-03-2018, a configuration generated by the UBCM-induced $\text{CReM}_\text{A}$ model and a configuration generated by the deterministic $\text{CReM}_\text{A}$ model for the same day. The latter one distributes weights more evenly than observed, hence underestimating disparity and letting the strength distribution become wider: this has an interesting consequence, i.e. letting the size of the core become larger than observed, under this model - it amounts at the $12\%$, the $27\%$ and the $21\%$ of the total number of nodes on the considered day - while still allowing the unevenness of the WEC distribution rise.}
\label{fig10}
\end{center}
\end{figure}

\paragraph{Disparity.} As anticipated in the paragraph introducing such a quantity, the disparity index of node $i$ reads $Y_i=1/k_i$ in case weights are equally distributed among the neighbors of node $i$. Figure \ref{fig9} shows the scatter plot of $Y_i$ as a function of $k_i$ (since it is plotted in a log-log scale, the function $y=-x$ becomes the trend signalling that weights are uniformly distributed among the neighbors of each node): generally speaking, many values lie above the $y=-x$ line, an evidence indicating that some kind of `excess concentration' of weight (in one or more links) is indeed present - a tendency which is particularly evident for nodes with smaller degree. 

Let us now compare the empirical disparity values with the predictions of the null models defined within our $\text{CReM}_\text{A}$ framework. To this aim, let us explicitly calculate the expected value of disparity, that reads

\begin{equation}
\langle Y_i\rangle\simeq\frac{\sum_{j(\neq i)=1}^N\langle w_{ij}^2\rangle}{\langle s_i^2\rangle}=\frac{\sum_{j(\neq i)=1}^N\left(\text{Var}[w_{ij}]+\langle w_{ij}\rangle^2\right)}{\text{Var}[s_i]+\langle s_i\rangle^2}=\frac{\text{Var}[s_i]+\sum_{j(\neq i)=1}^N\langle w_{ij}\rangle^2}{\text{Var}[s_i]+s_i^2},\quad\forall\:i
\end{equation}
where

\begin{equation}
\langle w_{ij}\rangle=\frac{p_{ij}}{\beta_i+\beta_j},\quad\forall\:i<j
\end{equation}
and

\begin{equation}
\text{Var}[s_i]=\sum_{j(\neq i)=1}^N\text{Var}[w_{ij}]=\sum_{j(\neq i)=1}^N\frac{p_{ij}}{(\beta_i+\beta_j)^2},\quad\forall\:i
\end{equation}
(naturally, for the the present analysis we have considered both the case $p_{ij}\equiv p_{ij}^\text{UBCM}$, $\forall\:i<j$ and the case $p_{ij}\equiv a_{ij}$, $\forall\:i<j$). As fig. \ref{fig9} shows, disparity is, generally speaking, in agreement with our benchmark models. However, the calculation of the percentage of nodes for which $Q(Y_i\geq Y_i^*)<0.05$, for our usual four snapshots, reveals it to be $0\%, 3.0\%, 9.1\%, 11\%$ for the UBCM-induced $\text{CReM}_\text{A}$ model and $0\%, 5.6\%, 15\%, 17\%$ for the deterministic $\text{CReM}_\text{A}$ model: in other words, the percentage of nodes whose empirical disparity is significantly larger than predicted by one of our two null models is rising throughout the entire BLN history. This evidence suggests that, as the BLN evolves, its vertices treat their neighbors less and less equally: indeed, they seem to place weights in a way that increasingly `favors' some of the links surrounding them - a result that remains true even when a null model constraining the entire topology of the BLN is employed\footnote{To be noticed that our null models also underestimate disparity values: however, the corresponding percentages (amounting at $3.2\%, 4.3\%, 5.8\%, 5.2\%$ for the UBCM-induced $\text{CReM}_\text{A}$ model and at $12.8\%, 10.8\%, 13.0\%, 13.6\%$ for the deterministic $\text{CReM}_\text{A}$ model, for our usual four snapshots), are typically lower and not increasing.}.

\paragraph{Centrality.} Let us now comment the results concerning the weighted centrality measures considered in the present work. As a general observation, the weighted cases are characterised by trends which are overall similar to the trends characterising the binary cases. As already observed for the purely binary BLN structure, the evolution of the Gini index for most centrality measures points out the latter ones to grow (strongly) unevenly distributed throughout the entire BLN history. While the rise of the Gini coefficient for the weighted degree, betweenness and eigenvector centrality measures suggests the appearance of nodes with `heavy' connections - further confirmed by the strength distribution, which is a fat-tailed one - likely crossed by many paths and well connected between themselves, the flat trend characterising the evolution of the closeness centrality confirms what has been already observed in the purely binary case, i.e. that the aforementioned `hubs' ensure the vast majority of nodes to be reachable (hence, to be close to each other) quite easily.

Let us now compare the empirical trends of our four centrality measures with the ones predicted by our two null models. Figure \ref{fig10} reveals that the UBCM-induced $\text{CReM}_\text{A}$ model tends to overestimate the values of the Gini index for the weighted degree and closeness centrality, i.e. the empirical weighted degree and closeness centrality measures are always significantly lower than their predicted counterparts. For what concerns the weighted betweenness centrality, instead, the percentage of snapshots for which $Q(G_\text{WBC}\geq G_\text{WBC}^*)<0.05$ amounts at $\simeq 87\%$, i.e. the UBCM-induced $\text{CReM}_\text{A}$ model significantly underestimates the weighted betweenness centrality for $\simeq 87\%$ of the total number of snapshots. Analogously, the same null model tends to underestimate the values of the Gini index for the weighted eigenvector centrality roughly one third of the times: in fact, the percentage of snapshots for which $Q(G_\text{WEC}\geq G_\text{WEC}^*)<0.05$ amounts at $\simeq 33\%$. Interestingly, the empirical WBC and WEC values are compatible with the predictions output by the UBCM-induced $\text{CReM}_\text{A}$ model, on the `remaining' snapshots.

\begin{figure}[t!]
\begin{center}
\includegraphics[width=\textwidth]{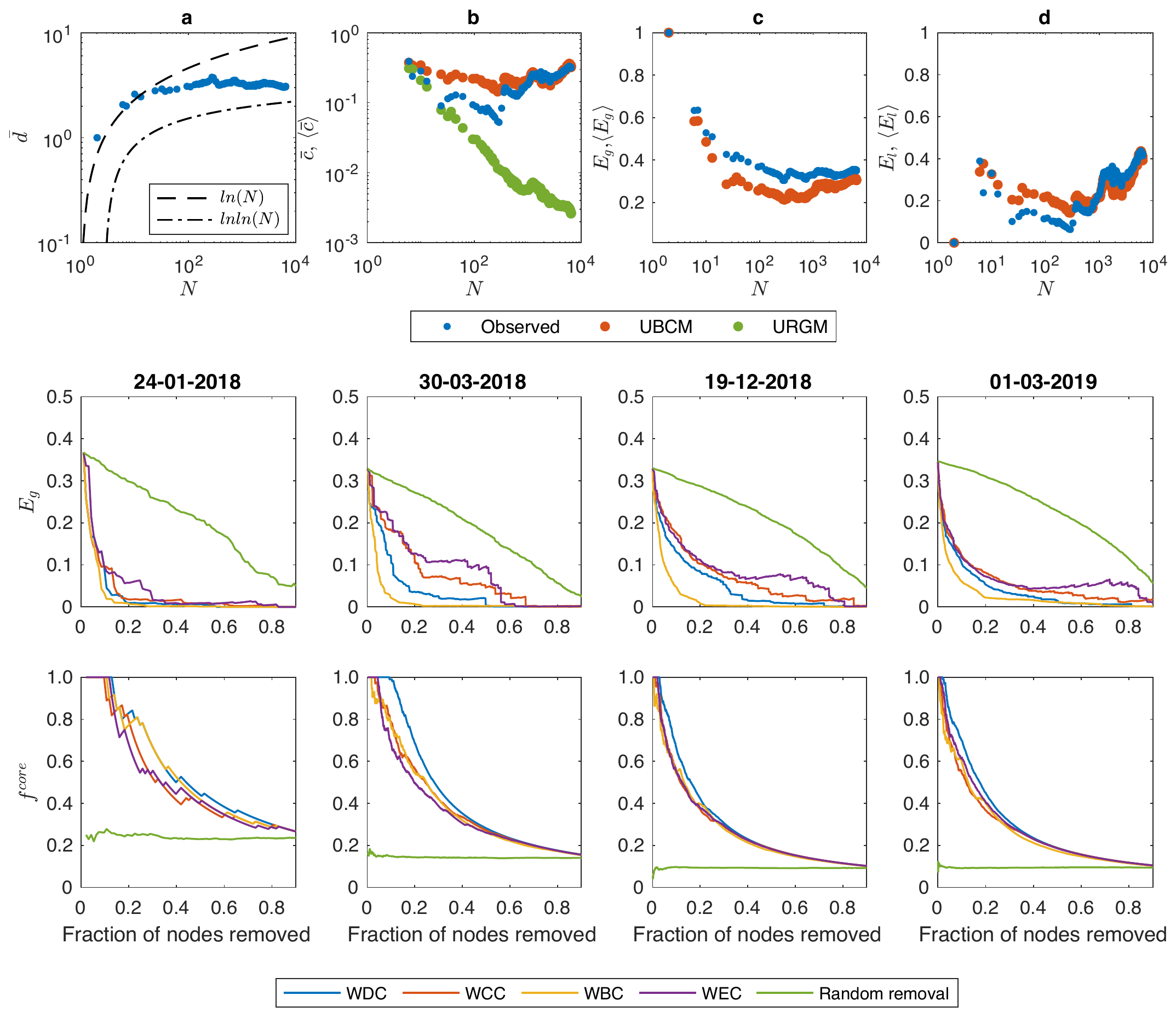}
\caption{Upper panels: (a) evolution of the BLN average path length $\bar{d}$ and of the functions $\ln N$ and $\ln\ln N$; (b) evolution of the empirical average clustering coefficient $\overline{\text{BCC}}$ (blue dots) and of its expected values $\langle\overline{\text{BCC}}\rangle$ under the URGM (green stars) and the UBCM (red squares); (c) evolution of the empirical global efficiency $E_g$ (blue dots) and of its expected values $\langle{E_g}\rangle$ under the UBCM (red squares); (d) evolution of the empirical local efficiency $E_l$ and of its expected values $\langle{E_l}\rangle$ under the UBCM (red squares). The BLN is indeed characterised by a small-world structure; moreover, while it has been always more-globally-efficient-than-expected under the UBCM, it has `recently' become also more-locally-efficient-than-expected under the same null model. Middle panels: evolution of the BLN global efficiency, for our usual four snapshots, when nodes are removed either randomly (green trend) or sequentially, after having been sorted in decreasing order of weighted degree (blue trend), closeness (red trend), betweenness (yellow) and eigenvector (purple) centrality. The trends above characterise a robust-yet-fragile architecture: robust against a random node removal but fragile against a targeted node removal (e.g. an attack). Bottom panels: percentage of core nodes found within the set of nodes removed according to one of the two aforementioned criteria.}
\label{fig11}
\end{center}
\end{figure}

The deterministic $\text{CReM}_\text{A}$ model, instead, performs slightly better in reproducing the centrality patterns characterising the BLN: in fact, while it still overestimates the Gini index for the weighted degree centrality, the percentage of snapshots for which $Q(G_\text{WCC}\geq G_\text{WCC}^*)<0.05$ amounts at $\simeq 96\%$, i.e. the deterministic $\text{CReM}_\text{A}$ model significantly underestimates the weighted closeness centrality for $\simeq 96\%$ of the total number of snapshots. For what concerns the weighted betweenness centrality, the percentage of snapshots for which $Q(G_\text{WBC}\geq G_\text{WBC}^*)<0.05$ amounts at $\simeq 50\%$, i.e. the deterministic $\text{CReM}_\text{A}$ model significantly underestimates the weighted betweenness centrality roughly half of the times. Lastly, for what concerns the weighted eigenvector centrality, the percentage of snapshots for which $Q(G_\text{WEC}\leq G_\text{WEC}^*)<0.05$ amounts at $\simeq 83\%$, i.e. the deterministic $\text{CReM}_\text{A}$ model significantly overestimates the weighted eigenvector centrality for $\simeq 83\%$ of the total number of snapshots. The deterministic $\text{CReM}_\text{A}$ model distributes weights more evenly than observed, hence underestimating disparity and letting the strength distribution become wider: this has an interesting consequence, i.e. letting the size of the core under this model become larger than observed - likely, because nodes with relatively low strength become, now, part of the core - while still allowing the unevenness of the WEC distribution rise.

Overall, these results point out a behavior that is not reproducible by just enforcing the degree and the strength sequences - irrespectively from the chosen index: in particular, the behavior of the weighted betweenness centrality points out that both null models - even if to a different extent - predict a more-even-than-observed structure.

\begin{figure}[t!]
\begin{center}
\includegraphics[width=\textwidth]{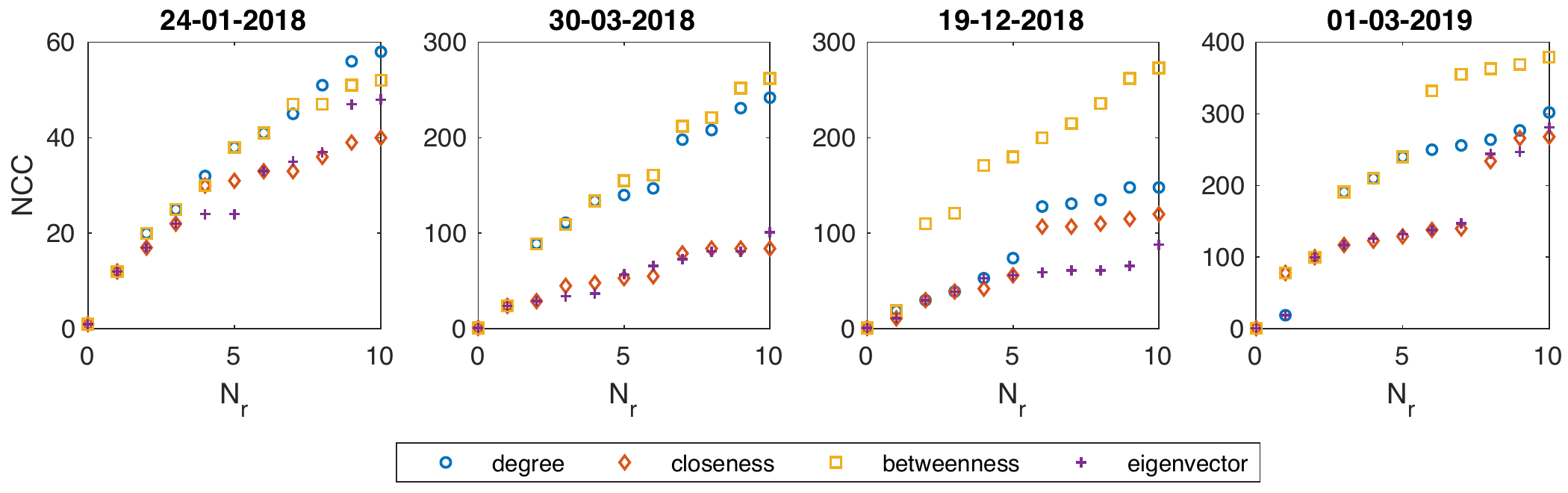}
\caption{Number of connected components (NCC) into which the BLN structure fragments upon removing nodes scattered versus the number of nodes ($N_r$ - ranked by degree, closeness, betweenness and eigenvector centrality) whose removal induces them.}
\label{fig12}
\end{center}
\end{figure}

\paragraph{Small-world -ness.} The evidence that the BLN structure is more-centralised-than-expected rises an interesting question, i.e. if the BLN is small-world or not. From a purely empirical perspective, answering this question amounts at checking the behavior of the average path length, $\overline{d}$, and that of the average clustering coefficient, $\overline{\text{BCC}}=\sum_i\text{BCC}_i/N$ \cite{watts1998collective,latora2001efficient,amara2011classes}.

Figure \ref{fig11} shows the results of these two analyses: while the evolution of $\overline{d}$ is described quite accurately by the function $\ln N$ during the first snapshots of the BLN history, its trend has progressively become more and more similar to the smoothest one characterising the function $\ln\ln N$ - which has reached the value $\simeq 3.5$ on the snapshot with $10^4$ nodes. For what concerns the average clustering coefficient, one needs to compare it with the value predicted by the URGM, i.e. the null model prescribing to link each pair of nodes with the same probability $p=2L/N(N-1)$: as fig. \ref{fig11} shows, the URGM significantly underestimates the average clustering coefficient throughout the entire BLN history; taken together, there results indicate that the BLN is indeed small-world. On the other hand, the UBCM overestimates $\overline{\text{BCC}}=\sum_i\text{BCC}_i/N$ during the first half of its history (for $\simeq 40\%$ of the total number of snapshots), thus signalling a tendency of our system to avoid closing paths among triples of nodes.

An alternative way of testing small-world -ness is that of checking the behavior of efficiency. Overall, the global efficiency amounts at $E_g\simeq 0.4$ and it is significantly underestimated by the UBCM throughout the entire history of the BLN. This indicates that the BLN exchanges information more-efficiently-than-predicted by a null model retaining only the information provided by degrees and can be a consequence of the presence of hubs crossed by many paths that shorten the topological distance between (any pair of) nodes. 

These results suggest that the BLN has progressively self-organized to `keep the overall distances low'. What about efficiency from a local point of view? For what concerns the local efficiency, the percentage of nodes for which $Q(E_l\geq E_l^*)<0.05$ amounts at $75\%$: hence, the UBCM significantly underestimates it for a large portion of the BLN snapshots - as evident from fig. \ref{fig11}, the most recent ones. As the local efficiency $E(\mathbf{G}_i)$ provides information about how efficient the communication between the first neighbors of node $i$ is, upon its removal, our results seem to indicate that the BLN is becoming more and more `fault tolerant' than its randomised counterpart (interestingly, it appeared to be much more fragile during the first half of its history). This result can be understood by imagining that a larger number of redundant connections has been established, among nodes, in the more recent snapshots of the BLN history - whence the rise of the average clustering coefficient as well.

As an additional exercise, let us inspect the evolution of the BLN global efficiency as nodes are removed either randomly or sequentially, after they have been sorted in decreasing order of weighted degree, closeness, betweenness and eigenvector centrality. The results of our exercise are shown in figs. \ref{fig11} and \ref{fig12}. The depicted trends of fig. \ref{fig11} are compatible with a \emph{robust-yet-fragile} architecture, i.e. a topological structure that is robust against a random removal of nodes but fragile against a targeted removal of nodes (e.g. an attack) - or, more correctly, more robust against a random node removal than against a targeted node removal: notice how steeper the decrease of $E_g$ is in the second case; moreover, removing nodes according to their WBC reduces the BLN global efficiency to the largest extent (for the vast majority of snapshots, larger than removing nodes according to their WDC, WCC and WEC). Figure \ref{fig12} sheds light on the reason why such an effect is observed: removing nodes induces a fragmentation of the BLN structure into many (disconnected) components.

Figure \ref{fig11} also shows that the nodes whose removal brings the most severe damages to the BLN are those belonging to the core (see the next paragraph), whose size shrinks from $\gtrsim 20\%$ to $\simeq 10\%$ of the total number of nodes.

\begin{figure}[t!]
\begin{center}
\includegraphics[width=\textwidth]{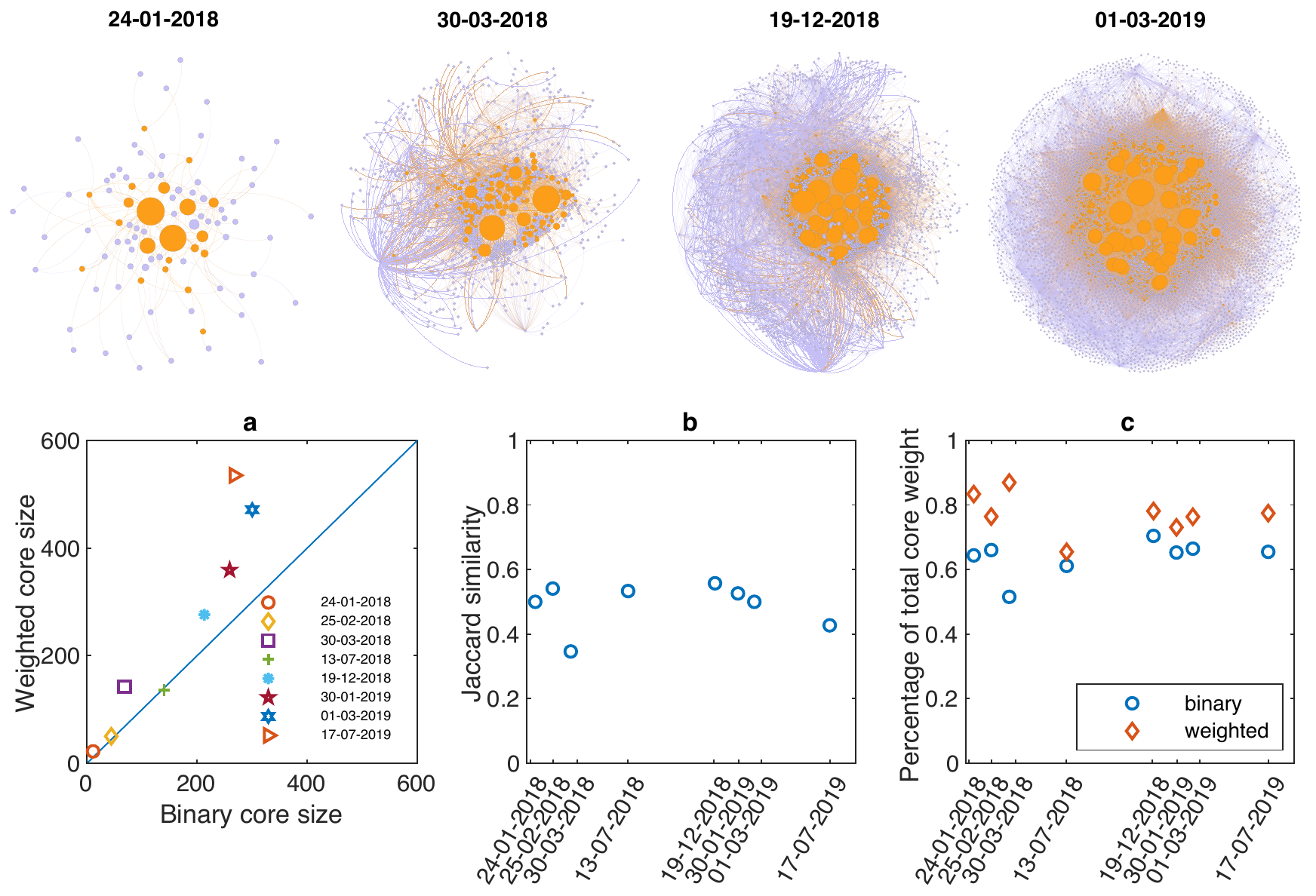}
\caption{Upper panels: core-periphery structure, as revealed by the weighted surprise, $\mathscr{W}_\sslash$, for our four usual snapshots (core nodes are colored in orange and periphery nodes are colored in purple); the size of the nodes is proportional to their strength: hence, the nodes constituting the core of the network are precisely those with a larger strength. Bottom panels: (a) values of the size of the binary core scattered versus the values of the size of the weighted core; (b) evolution of (the percentage of) the overlap between the set of nodes belonging to the binary core and the set of nodes belonging to the weighted one, estimated via the Jaccard similarity, on a selected subset of snapshots of the BLN; evolution of the percentage of the total network weight embodied by `core connections', amounting at $\simeq 68\%$ in the binary case (blue circles) and at $\simeq 77\%$ in the weighted one (red diamonds).}
\label{fig13}
\end{center}
\end{figure}

\paragraph{Core-periphery detection.} The result concerning the underestimation, by our null models, of the Gini index for the weighted eigenvector centrality - i.e. the presence of well connected nodes which, in turn, are also well connected among them - lets us suppose the BLN to be characterised by a statistically significant core-periphery structure: here, however, we are interested in revealing the presence of a weighted core-periphery structure, i.e. a kind of mesoscale organisation where core nodes are the ones sharing the `heaviest' connections - and not just those with `many' connections.

To this aim, we adopt a recently proposed approach, based upon the surprise formalism. In particular, we consider the evolution of the weighted bimodular surprise, $\mathscr{W}_\sslash$, across the entire BLN history: it reveals that the statistical significance of the recovered core-periphery structure increases, a result leading to the conclusion that the description of the BLN structure provided by such a model becomes more and more accurate as the network evolves. As an example, fig. \ref{fig12} shows the detected core-periphery structure on the snapshots depicted in the same figure: the nodes identified as belonging to the core and to the periphery are, respectively, colored in orange and purple. Notice also that we have drawn the node size proportionally to the node strength: hence, larger nodes, i.e. the ones sharing the `heaviest' connections, are precisely those constituting the core of the network.

First, let us check the correspondence between the nodes in the core (whose size will be indicated as $N_{core}$) and vertices with large weighted eigenvector centrality by ranking the nodes in decreasing order of WEC and checking the percentage of top $N_{core}$ nodes that also belong to the core: it amounts at $56\%, 60\%, 57\%, 62\%$, for our usual four snapshots. Then, let us compare the composition of the purely binary core - detected in \cite{lin2020lightning} - with that of the weighted core. As fig. \ref{fig12} shows, a nice correspondence between the size of binary core and that of the weighted one indeed exists although, from a certain moment of the BLN history on, the binary core seems to `grow slower' than the weighted one which, instead, enlarges to reach a size of $\simeq 600$ nodes: this further confirms that the nodes with a `large' strength, revealed by surprise as the most central ones, do not necessarily coincide with those having a `large' degree.

The evolution of (the percentage of) the overlap between the set of nodes belonging to the binary core and the set of nodes belonging to the weighted one further confirms that the two sets do not coincide perfectly, although the Jaccard similarity steadily points out a $\simeq 60\%$ of overlap: in other words, $60\%$ of the nodes belong to both cores - likely, those hubs whose degree \emph{and} strength are large enough to justify their coreness in both senses; similarly, the percentage of the total network weight embodied by `core connections' amounts at $\simeq 68\%$ in the binary case and at $\simeq 77\%$ in the weighted one.

\section{Discussion}

The analysis of the binary BLN structure carried out in \cite{lin2020lightning} has revealed a system whose topology has become increasingly characterised by star-like structures, whose centers are constituted by `hubs' to which many nodes having a (much) small(er) degree, in turn, attach. Such a structure - whose disassortativity is confirmed by scattering the ANND values versus the degrees - could explain the more-than-expected level of unevenness characterizing the betweenness and the eigenvector centralisation indices, suggesting them to be due to the emergence of channel-switching nodes - apparently, an unavoidable consequence of the way BLN is designed: on the one hand, as longer routes are more expensive, any two BLN users will search for a short(est) path; on the other, nodes have the incentive to become as central as possible, in order to maximize the transaction fees they may earn.

The tendency to centralisation is observable even when considering weighted quantities, as the percentage of nodes whose connections embody the $51\%$ of the total weight progressively reduces and the Gini coefficient of several (weighted) centrality measures steadily increases throughout the entire BLN history. This clearly points out the co-existence of nodes playing deeply different `structural' roles, with `many' peripheral vertices co-existing with `few' core ones; if, on the one hand, this structure allows the global efficiency to achieve a large value (i.e. hubs facilitate the global exchange of information, being at the origin of another structural BLN peculiarity, i.e. its small-world -ness), on the other it highlights the tendency of the BLN architecture to become increasingly `less distributed', a process having the undesirable consequence of making it increasingly fragile towards failures and attacks.

Distinguishing between the two is crucial, in order to properly understand the BLN robustness to `damages'. While resilience towards failures can be tested by looking at how the global efficiency `reacts' to random node removal, resilience towards attacks can, instead, be quantified by implementing targeted removals of the `most important' nodes. To this aim, we have ranked nodes in decreasing order of weighted degree, closeness, betweenness and eigenvector centrality and removed them, sequentially: the global efficiency drops rapidly after few (core) nodes are deleted - in fact, for almost all snapshots, removing just \emph{one top node} (according to any of the aforementioned criteria) is enough to disconnect the graph. Moreover, since top nodes are likely to be part of the core - whose size shrinks from $\gtrsim 20\%$ to $\simeq 10\%$ of the total number of nodes - our results indicate that the vertices belonging to it are precisely those whose removal causes the major structural damages. Random failures, instead, cause the decrease of $E_g$ to be much less steep: taken together, the results above seem to indicate that the BLN topology is an example of \emph{robust-yet-fragile} architecture, i.e. a structure that is robust against a random node removal but fragile against a targeted node removal (e.g. an attack).

\section{Authors contributions}

Jian-Hong Lin and Emiliano Marchese performed the analysis. Claudio J. Tessone and Tiziano Squartini designed the research. All authors wrote, reviewed and approved the manuscript.

\section{Acknowledgements}

J.-H. L. and C.J.T. acknowledge financial support from the Swiss National Science Foundation grant no. $\#200021\_182659$. J.-H. L. also acknowledges financial support from the China Scholarship Council. T.S. acknowledges financial support from the EU H2020 project `SoBigData++', grant n. 871042.

\bibliography{Bibliography}

\begin{thebibliography}{10}
\expandafter\ifx\csname url\endcsname\relax
  \def\url#1{\texttt{#1}}\fi
\expandafter\ifx\csname urlprefix\endcsname\relax\def\urlprefix{URL }\fi
\providecommand{\bibinfo}[2]{#2}
\providecommand{\eprint}[2][]{\url{#2}}

\bibitem{poon2016bitcoin}
\bibinfo{author}{Poon, J.} \& \bibinfo{author}{Dryja, T.}
\newblock \bibinfo{title}{The bitcoin lightning network: Scalable off-chain
  instant payments} (\bibinfo{year}{2016}).

\bibitem{branzei2017charge}
\bibinfo{author}{Br{\^a}nzei, S.}, \bibinfo{author}{Segal-Halevi, E.} \&
  \bibinfo{author}{Zohar, A.}
\newblock \bibinfo{title}{How to charge lightning}.
\newblock \emph{\bibinfo{journal}{arXiv preprint arXiv:1712.10222}}
  (\bibinfo{year}{2017}).

\bibitem{beres2021cryptoeconomic}
\bibinfo{author}{B{\'e}res, F.}, \bibinfo{author}{Seres, I.~A.},
  \bibinfo{author}{Bencz{\'u}r, A.} \emph{et~al.}
\newblock \bibinfo{title}{A cryptoeconomic traffic analysis of bitcoin’s
  lightning network}.
\newblock \emph{\bibinfo{journal}{CRYPTOECONOMIC SYSTEMS}}
  \textbf{\bibinfo{volume}{1}}, \bibinfo{pages}{1--46} (\bibinfo{year}{2021}).

\bibitem{lee2020robustness}
\bibinfo{author}{Lee, S.} \& \bibinfo{author}{Kim, H.}
\newblock \bibinfo{title}{On the robustness of lightning network in bitcoin}.
\newblock \emph{\bibinfo{journal}{Pervasive and Mobile Computing}}
  \textbf{\bibinfo{volume}{61}}, \bibinfo{pages}{101108}
  (\bibinfo{year}{2020}).

\bibitem{lin2020lightning}
\bibinfo{author}{Lin, J.-H.}, \bibinfo{author}{Primicerio, K.},
  \bibinfo{author}{Squartini, T.}, \bibinfo{author}{Decker, C.} \&
  \bibinfo{author}{Tessone, C.~J.}
\newblock \bibinfo{title}{Lightning network: a second path towards
  centralisation of the bitcoin economy}.
\newblock \emph{\bibinfo{journal}{New Journal of Physics}}
  \textbf{\bibinfo{volume}{22}}, \bibinfo{pages}{083022}
  (\bibinfo{year}{2020}).

\bibitem{martinazzi2020evolving}
\bibinfo{author}{Martinazzi, S.} \& \bibinfo{author}{Flori, A.}
\newblock \bibinfo{title}{The evolving topology of the lightning network:
  Centralization, efficiency, robustness, synchronization, and anonymity}.
\newblock \emph{\bibinfo{journal}{PLoS ONE}} \textbf{\bibinfo{volume}{15}},
  \bibinfo{pages}{e0225966} (\bibinfo{year}{2020}).

\bibitem{motamed2019quantitative}
\bibinfo{author}{Motamed, A.~P.} \& \bibinfo{author}{Bahrak, B.}
\newblock \bibinfo{title}{Quantitative analysis of cryptocurrencies transaction
  graph}.
\newblock \emph{\bibinfo{journal}{Applied Network Science}}
  \textbf{\bibinfo{volume}{4}}, \bibinfo{pages}{1--21} (\bibinfo{year}{2019}).

\bibitem{campajola2022evolution}
\bibinfo{author}{Campajola, C.} \emph{et~al.}
\newblock \bibinfo{title}{The evolution of centralisation on cryptocurrency
  platforms}.
\newblock \emph{\bibinfo{journal}{arXiv preprint arXiv:2206.05081}}
  (\bibinfo{year}{2022}).

\bibitem{guo2019measurement}
\bibinfo{author}{Guo, Y.}, \bibinfo{author}{Tong, J.} \& \bibinfo{author}{Feng,
  C.}
\newblock \bibinfo{title}{A measurement study of bitcoin lightning network}.
\newblock In \emph{\bibinfo{booktitle}{2019 IEEE International Conference on
  Blockchain (Blockchain)}}, \bibinfo{pages}{202--211}
  (\bibinfo{organization}{IEEE}, \bibinfo{year}{2019}).

\bibitem{rohrer2019discharged}
\bibinfo{author}{Rohrer, E.}, \bibinfo{author}{Malliaris, J.} \&
  \bibinfo{author}{Tschorsch, F.}
\newblock \bibinfo{title}{Discharged payment channels: Quantifying the
  lightning network's resilience to topology-based attacks}.
\newblock In \emph{\bibinfo{booktitle}{2019 IEEE European Symposium on Security
  and Privacy Workshops (EuroS\&PW)}}, \bibinfo{pages}{347--356}
  (\bibinfo{organization}{IEEE}, \bibinfo{year}{2019}).

\bibitem{mizrahi2021congestion}
\bibinfo{author}{Mizrahi, A.} \& \bibinfo{author}{Zohar, A.}
\newblock \bibinfo{title}{Congestion attacks in payment channel networks}.
\newblock In \emph{\bibinfo{booktitle}{International Conference on Financial
  Cryptography and Data Security}}, \bibinfo{pages}{170--188}
  (\bibinfo{organization}{Springer}, \bibinfo{year}{2021}).

\bibitem{conoscenti2019hubs}
\bibinfo{author}{Conoscenti, M.}, \bibinfo{author}{Vetr{\`o}, A.} \&
  \bibinfo{author}{De~Martin, J.~C.}
\newblock \bibinfo{title}{Hubs, rebalancing and service providers in the
  lightning network}.
\newblock \emph{\bibinfo{journal}{IEEE Access}} \textbf{\bibinfo{volume}{7}},
  \bibinfo{pages}{132828--132840} (\bibinfo{year}{2019}).

\bibitem{bartolucci2020percolation}
\bibinfo{author}{Bartolucci, S.}, \bibinfo{author}{Caccioli, F.} \&
  \bibinfo{author}{Vivo, P.}
\newblock \bibinfo{title}{A percolation model for the emergence of the bitcoin
  lightning network}.
\newblock \emph{\bibinfo{journal}{Scientific Reports}}
  \textbf{\bibinfo{volume}{10}}, \bibinfo{pages}{1--14} (\bibinfo{year}{2020}).

\bibitem{javarone2018bitcoin}
\bibinfo{author}{Javarone, M.~A.} \& \bibinfo{author}{Wright, C.~S.}
\newblock \bibinfo{title}{From bitcoin to bitcoin cash: a network analysis}.
\newblock In \emph{\bibinfo{booktitle}{Proceedings of the 1st Workshop on
  Cryptocurrencies and Blockchains for Distributed Systems}},
  \bibinfo{pages}{77--81} (\bibinfo{year}{2018}).

\bibitem{wu2021analysis}
\bibinfo{author}{Wu, J.}, \bibinfo{author}{Liu, J.}, \bibinfo{author}{Zhao, Y.}
  \& \bibinfo{author}{Zheng, Z.}
\newblock \bibinfo{title}{Analysis of cryptocurrency transactions from a
  network perspective: An overview}.
\newblock \emph{\bibinfo{journal}{Journal of Network and Computer
  Applications}} \textbf{\bibinfo{volume}{190}}, \bibinfo{pages}{103139}
  (\bibinfo{year}{2021}).

\bibitem{lngossip}
\bibinfo{author}{Decker, C.}
\newblock \bibinfo{title}{Lightning network research; topology datasets}.
\newblock \bibinfo{howpublished}{\url{https://github.com/lnresearch/topology}}.
\newblock \bibinfo{note}{Accessed: 2020-10-01}.

\bibitem{newman2003structure}
\bibinfo{author}{Newman, M.}
\newblock \bibinfo{title}{The structure and function of complex networks}.
\newblock \emph{\bibinfo{journal}{SIAM Review}} \textbf{\bibinfo{volume}{45}},
  \bibinfo{pages}{167--256} (\bibinfo{year}{2003}).

\bibitem{newman2006structure}
\bibinfo{author}{Newman, M.}, \bibinfo{author}{Barab{\'a}si, A.-L.} \&
  \bibinfo{author}{Watts, D.~J.}
\newblock \emph{\bibinfo{title}{The structure and dynamics of networks.}}
  (\bibinfo{publisher}{Princeton university press}, \bibinfo{year}{2006}).

\bibitem{srinivasan2017quantifying}
\bibinfo{author}{Srinivasan, B.~S.} \& \bibinfo{author}{Lee, L.}
\newblock \bibinfo{title}{Quantifying decentralization}.
\newblock
  \emph{\bibinfo{journal}{https://news.earn.com/quantifying-decentralization-e39db233c28e}}
   (\bibinfo{year}{2017}).

\bibitem{bonacich1987power}
\bibinfo{author}{Bonacich, P.}
\newblock \bibinfo{title}{Power and centrality: A family of measures}.
\newblock \emph{\bibinfo{journal}{American Journal of Sociology}}
  \textbf{\bibinfo{volume}{92}}, \bibinfo{pages}{1170--1182}
  (\bibinfo{year}{1987}).

\bibitem{borgatti2005centrality}
\bibinfo{author}{Borgatti, S.~P.}
\newblock \bibinfo{title}{Centrality and network flow}.
\newblock \emph{\bibinfo{journal}{Social Networks}}
  \textbf{\bibinfo{volume}{27}}, \bibinfo{pages}{55--71}
  (\bibinfo{year}{2005}).

\bibitem{newman2018networks}
\bibinfo{author}{Newman, M.}
\newblock \emph{\bibinfo{title}{Networks. An introduction}}
  (\bibinfo{publisher}{Oxford university press}, \bibinfo{year}{2018}).

\bibitem{rodrigues2019network}
\bibinfo{author}{Rodrigues, F.~A.}
\newblock \bibinfo{title}{Network centrality: an introduction}.
\newblock In \emph{\bibinfo{booktitle}{A mathematical modeling approach from
  nonlinear dynamics to complex systems}}, \bibinfo{pages}{177--196}
  (\bibinfo{publisher}{Springer}, \bibinfo{year}{2019}).

\bibitem{newman2005measure}
\bibinfo{author}{Newman, M.}
\newblock \bibinfo{title}{A measure of betweenness centrality based on random
  walks}.
\newblock \emph{\bibinfo{journal}{Social Networks}}
  \textbf{\bibinfo{volume}{27}}, \bibinfo{pages}{39--54}
  (\bibinfo{year}{2005}).

\bibitem{pfitzner2013betweenness}
\bibinfo{author}{Pfitzner, R.}, \bibinfo{author}{Scholtes, I.},
  \bibinfo{author}{Garas, A.}, \bibinfo{author}{Tessone, C.~J.} \&
  \bibinfo{author}{Schweitzer, F.}
\newblock \bibinfo{title}{Betweenness preference: Quantifying correlations in
  the topological dynamics of temporal networks}.
\newblock \emph{\bibinfo{journal}{Physical Review Letters}}
  \textbf{\bibinfo{volume}{110}}, \bibinfo{pages}{198701}
  (\bibinfo{year}{2013}).

\bibitem{bonacich2007some}
\bibinfo{author}{Bonacich, P.}
\newblock \bibinfo{title}{Some unique properties of eigenvector centrality}.
\newblock \emph{\bibinfo{journal}{Social Networks}}
  \textbf{\bibinfo{volume}{29}}, \bibinfo{pages}{555--564}
  (\bibinfo{year}{2007}).

\bibitem{liu2016locating}
\bibinfo{author}{Liu, J.-G.}, \bibinfo{author}{Lin, J.-H.},
  \bibinfo{author}{Guo, Q.} \& \bibinfo{author}{Zhou, T.}
\newblock \bibinfo{title}{Locating influential nodes via dynamics-sensitive
  centrality}.
\newblock \emph{\bibinfo{journal}{Scientific Reports}}
  \textbf{\bibinfo{volume}{6}}, \bibinfo{pages}{21380} (\bibinfo{year}{2016}).

\bibitem{morgan1962anatomy}
\bibinfo{author}{Morgan, J.}
\newblock \bibinfo{title}{The anatomy of income distribution}.
\newblock \emph{\bibinfo{journal}{The Review of Economics and Statistics}}
  \bibinfo{pages}{270--283} (\bibinfo{year}{1962}).

\bibitem{crucitti2006centrality}
\bibinfo{author}{Crucitti, P.}, \bibinfo{author}{Latora, V.} \&
  \bibinfo{author}{Porta, S.}
\newblock \bibinfo{title}{Centrality measures in spatial networks of urban
  streets}.
\newblock \emph{\bibinfo{journal}{Physical Review E}}
  \textbf{\bibinfo{volume}{73}}, \bibinfo{pages}{036125}
  (\bibinfo{year}{2006}).

\bibitem{clauset2009power}
\bibinfo{author}{Clauset, A.}, \bibinfo{author}{Shalizi, C.~R.} \&
  \bibinfo{author}{Newman, M.~E.}
\newblock \bibinfo{title}{Power-law distributions in empirical data}.
\newblock \emph{\bibinfo{journal}{SIAM review}} \textbf{\bibinfo{volume}{51}},
  \bibinfo{pages}{661--703} (\bibinfo{year}{2009}).

\bibitem{watts1998collective}
\bibinfo{author}{Watts, D.~J.} \& \bibinfo{author}{Strogatz, S.~H.}
\newblock \bibinfo{title}{Collective dynamics of ‘small-world’networks}.
\newblock \emph{\bibinfo{journal}{Nature}} \textbf{\bibinfo{volume}{393}},
  \bibinfo{pages}{440--442} (\bibinfo{year}{1998}).

\bibitem{latora2001efficient}
\bibinfo{author}{Latora, V.} \& \bibinfo{author}{Marchiori, M.}
\newblock \bibinfo{title}{Efficient behavior of small-world networks}.
\newblock \emph{\bibinfo{journal}{Physical Review Letters}}
  \textbf{\bibinfo{volume}{87}}, \bibinfo{pages}{198701}
  (\bibinfo{year}{2001}).

\bibitem{amara2011classes}
\bibinfo{author}{Amara, L.}, \bibinfo{author}{Scala, A.},
  \bibinfo{author}{Barthelemy, M.} \& \bibinfo{author}{Stanley, H.~E.}
\newblock \bibinfo{title}{Classes of small-world networks}.
\newblock In \emph{\bibinfo{booktitle}{The Structure and Dynamics of
  Networks}}, \bibinfo{pages}{207--210} (\bibinfo{publisher}{Princeton
  University Press}, \bibinfo{year}{2011}).

\bibitem{marchese2022detecting}
\bibinfo{author}{Marchese, E.}, \bibinfo{author}{Caldarelli, G.} \&
  \bibinfo{author}{Squartini, T.}
\newblock \bibinfo{title}{Detecting mesoscale structures by surprise}.
\newblock \emph{\bibinfo{journal}{Communications Physics}}
  \textbf{\bibinfo{volume}{5}}, \bibinfo{pages}{1--16} (\bibinfo{year}{2022}).

\bibitem{dejeude2019detecting}
\bibinfo{author}{van Lidth~de Jeude, J.}, \bibinfo{author}{Caldarelli, G.} \&
  \bibinfo{author}{Squartini, T.}
\newblock \bibinfo{title}{Detecting core-periphery structures by surprise}.
\newblock \emph{\bibinfo{journal}{Europhysics Letters}}
  \textbf{\bibinfo{volume}{125}}, \bibinfo{pages}{68001}
  (\bibinfo{year}{2019}).

\bibitem{parisi2020faster}
\bibinfo{author}{Parisi, F.}, \bibinfo{author}{Squartini, T.} \&
  \bibinfo{author}{Garlaschelli, D.}
\newblock \bibinfo{title}{A faster horse on a safer trail: generalized
  inference for the efficient reconstruction of weighted networks}.
\newblock \emph{\bibinfo{journal}{New Journal of Physics}}
  \textbf{\bibinfo{volume}{22}}, \bibinfo{pages}{053053}
  (\bibinfo{year}{2020}).

\bibitem{vallarano2021fast}
\bibinfo{author}{Vallarano, N.} \emph{et~al.}
\newblock \bibinfo{title}{Fast and scalable likelihood maximization for
  exponential random graph models with local constraints}.
\newblock \emph{\bibinfo{journal}{Scientific Reports}}
  \textbf{\bibinfo{volume}{11}}, \bibinfo{pages}{1--33} (\bibinfo{year}{2021}).

\end{thebibliography}

\end{document}